\documentclass{aa}  

\usepackage{graphicx}
\usepackage{txfonts}
\usepackage{natbib}
\usepackage{subcaption}
\usepackage{threeparttable}
\usepackage{color}
\usepackage[colorlinks=true,linkcolor=blue]{hyperref}

\usepackage{url}
\usepackage{tabularx}

\usepackage[normalem]{ulem}

\begin{document} 
\defcitealias{NK13}{N13}

\title{How much metals did the first stars provide to the ultra-faint dwarfs?}

   \author{Mahsa Sanati\inst{\ref{epfl}},
           Fabien Jeanquartier\inst{\ref{epfl}}, 
           Yves Revaz\inst{\ref{epfl}}, 
           Pascale Jablonka\inst{\ref{epfl}, \ref{Observatoire de Paris}}
          }

   \institute{Institute of Physics, Laboratory of Astrophysics, \'{E}cole Polytechnique F\'{e}d\'{e}rale de Lausanne (EPFL), 1290 Sauverny, Switzerland \label{epfl}\\
        \email{mahsa.sanati@epfl.ch}
    \and GEPI, CNRS UMR 8111, Observatoire de Paris, PSL University, 92125 Meudon, Cedex, France \label{Observatoire de Paris}
             }

   \date{Received: XXXX; accepted: YYYY}
 
  \abstract
    { 
    Numerical simulations of dwarf galaxies have so far failed to reproduce the observed metallicity-luminosity relation, down to the regime of ultra-faint dwarfs (UFDs). 
    We address this issue by exploring how the first generations of metal-free stars (Pop III) could help increase the mean metallicity ($[\rm{Fe/H}]$) of those small and faint galaxies.
    We run zoom-in chemo-dynamical simulations of nineteen halos extracted from a $\Lambda$CDM cosmological box and follow down to redshift $z=0$. Models are validated not only on the basis of galaxy global properties, but also on the detailed investigation of the stellar abundance ratios ([$\alpha$/Fe]).
    We identify the necessary conditions for the formation of first stars in mini-halos and derive constraints on the metal ejection schemes.
    The impact of Pop III stars on the final metallicity of UFDs is evaluated by considering different IMFs, the influence of pair-instability supernovae (PISNe) and their energetic feedback, as well as the metallicity threshold that marks the transition from the first massive stars to the formation of low-mass long-lived stars.
    The inclusion of Pop III stars with masses below $140\,\mathrm{M}_{\odot}$, and a standard IMF slope of $-1.3$
    does increase the global metallicity of UFDs, though insufficient to resolve the tension with observations.
    PISNe with progenitor masses above $140\,\mathrm{M}_{\odot}$ do allow to further increase the metal content of UFDs. However, as PISNe are very rare and sometimes absent in the faintest UFDs, they have a limited impact on the global faint end of the metallicity-luminosity relation.  
    Despite a limited number of spectroscopically confirmed members in UFDs, that make the metallicity distribution of some UFDs uncertain,
    our analysis reveals that this is essentially the metal-rich tail that is missing in the models.
    The remaining challenges are thus both observational and numerical:    
    (i) to extend high resolution spectroscopy data samples and confirm the mean metallicity of the faintest UFDs,
    (ii) to explain the presence of chemically enriched stars in galaxies with very short star formation histories.
    
   }


   \titlerunning{Ultra-faint dwarfs - enrichment from first stars}   
   \authorrunning{M. Sanati et al}

   \maketitle

%

\section{Introduction}\label{sec:intro}


Ultra-faint dwarf galaxies (UFDs) are the faintest galaxies known, with V-band luminosities fainter 
than $L_\mathrm{V}$=$10^5\,L_{\odot}$, $M_{\rm{V}} < -7.7$ \citep[see][for a recent review]{Simon_2019}.
Some of them are as faint as few hundred solar luminosities, and may be as compact as faint 
globular clusters.
UFDs are also the most dark matter dominated galactic systems 
and as such they constitute fundamental 
probes of the cosmological model \citep[][]{Bullock2017}.

Semi-analytical and hydro-dynamical numerical simulations showed that UFDs are compatible with the first galaxies formed in mini-halos before the epoch of reionization  \citep[e.g.,][]{Ricotti_2005,Wyithe_2006,salvadori2009,2009MNRAS.395L...6S,bovill2009,Wheeler_2019,Rodriguez2019}.  These 
metal poor systems \citep{munoz2006,martin2007,simon2007,kirby2008}
have indeed formed the bulk of their stellar mass  ($\ge$ 80\%) by z $\sim$ 6 as revealed by deep ground- and space-based color-magnitude diagrams \citep{okamoto2012,brown2014, sacchi2021, gallart2021}.
Therefore, their properties offer a unique insight on the physics at work in the early Universe.


One might think that these small systems, with very short star formation histories, would be the product of simple physical processes. This is not the case. They do challenge numerical simulations, which struggle to reproduce their properties. Their small size, low mass and shallow gravitational potential makes them extremely sensitive to the numerical implementation of the physical processes; they require high resolution and they are critically sensitive to a correct and well controlled treatment of the first generations of stars, whose chemical signatures should be more salient than in more massive systems \citep[e.g., ][]{ji2015}.

Along this line, one of the most striking obstacles faced by numerical simulations 
is the reproduction of the observed luminosity-metallicity relation in UFDs. As described in detail in Sect.~\ref{sec:challenge}, all hydro-dynamical models predict systematically too low a metallicity at a given luminosity, compared to the observations, or in other words, too steep a slope of the $[\rm{Fe/H}]$-$\mathrm{L}_\mathrm{V}$ relation \citep{jeon2017,Maccio2017,Escala2018,Wheeler_2019,agertz2020, applebaum2020, jeon2021, prgomet2022}.
The question arises that whether the models are missing an important source of metals in the stellar populations so far 
and/or whether the nucleosynthesis products are too diluted in the interstellar medium (ISM) after the supernovae explosions.

The pair-instability supernovae (PISNe), first introduced by \cite{barakat1967}, are, in theory, an outstanding source of metals, hence worth considering in the context of UDFs.  
These supernovae result from an instability induced by the production of free electron-positron pairs
in metal-free stars with masses larger than $140\,\rm{M_\odot}$. This instability leads to a complete disruption of the star, releasing a large amount of heavy elements. 
Since UFDs cease forming stars shortly after the reionisation, PISNe could significantly influence their chemical evolution as part of the first (population III, hereafter Pop III) stars, pre-enriching the ISM. The influence of Pop III stars in UFDs has so far hardly been investigated. 
They are either included in the semi-analytical approach of \citet{salvadori2009}, or in the hydro-dynamical simulations of \citet{jeon2017}, where their specific influence has not been documented. 

No metal-free stars have yet been observed, suggesting that Pop III stars were sufficiently massive to either collapse into black holes or explode as a supernovae. In the latter case, they could leave their nucleosynthesis imprints in the next generations of low-mass long-lived stars (population II, hereafter Pop II) that are observed today.
However, the physical conditions under which they are formed and their detailed properties are still unsettled. This leaves wide open questions about the time frame 
of their formation 
in the cosmic history \citep[][]{2004ARA&A..42...79B, 2012PTEP.2012aA305Y, 2012ApJ...755...72H, 2016MNRAS.462.3591M}, the mass of the halos in which they were hosted \citep[see for example][]{skinner2020,schauer2020}, as well as their initial mass function (IMF) \citep[see for example][]{2008ApJ...682L...1M,2009Sci...325..601T,2010AIPC.1294..289S,2011Sci...331.1040C,2012RvMP...84...25M}. 
Chemo-dynamical numerical simulations of UFDs can test different Pop III scenarios and verify
their compatibility with the stellar abundance ratios observed in Pop II stars.

The goal of this paper is to precisely track down the conditions under which the global metal content of the model UFDs, traced by their iron abundance, can be increased at fixed stellar mass/luminosity. This involves investigating i) the physical conditions of star formation, ii) the possible strong Fe contributors among the very first stars, iii) the IMF of the first stars, iv) how metals are released and mixed in the ISM, as well as v) the impact of the supernovae feedback energy. Just as in our previous works, we discriminate between different scenarios not only on grounds of global galaxy properties (e.g., <$[\rm{Fe/H}]$>, luminosities, mass) but also on their detailed stellar abundance ratios,  metallicity distributions and velocity dispersions. 

The paper is organized as followed:
In Sec.~\ref{sec:challenge}, we review in details the discrepancy between the observed and simulated luminosity-metallicity relations, and the different scenarios that have been suggested to address this issue. In Sec.~\ref{sec:simulations}, we present our numerical methods, introduce in particular our  treatment of the Pop III stars, and describe the entire set of simulations. 
In Sec.~\ref{sec:results}, we discuss the
conditions from the formation of high mass to low mass stars, the IMF of Pop III, their explosion energy,
as well as their impact on the relation between metallicity and luminosity, with and without PISNe.
We also compare the $\alpha$-abundance ratios as well as the metallicity distribution function in models and observations.
Our conclusions are presented in Sec.~\ref{sec:conclusions}.



\section{The challenge of the luminosity-metallicity relation}\label{sec:challenge}


Similarly to more massive galaxies, dwarf galaxies fall on the scaling relations. 
Among those, the metallicity-luminosity
relation has been observed first by \citet{lequeux1979} for irregular and blue compact galaxies and subsequently confirmed by many authors 
extending both to dwarf galaxies or more distant spirals 
\citep[see for example][]{Skillman1989,garnett2002,tremonti2004}.
Basically, dwarf galaxies brighter than $L_{\rm{V}} > 10^5\,\rm{L_\odot}$ synthesise a larger quantity of metals which are further locked in the next generation of stars, leading to the observed metal-rich stellar systems. This is relatively well reproduced by the models.
However, in the faint (UFD) regime, the observed dispersion in metallicity at fixed luminosity increases \citep{Simon_2019}, and the slope of the observed relation is less steep than the  models, with lower  $\rm{[Fe/H]}$ at a given luminosity. 
This common feature is obtained  despite a variety of numerical schemes and  stellar feedback recipes that may influence the final metallicity \citep{agertz2020}. 
While this systematic deviation drew attention only recently, in particular 
with the recent improvements in numerical resolution, it has been already mentioned 
in past studies \citep[][see their Fig.~3]{bovill2009}. 

\begin{figure}[h]
  \centering
  \includegraphics[width=0.49\textwidth]{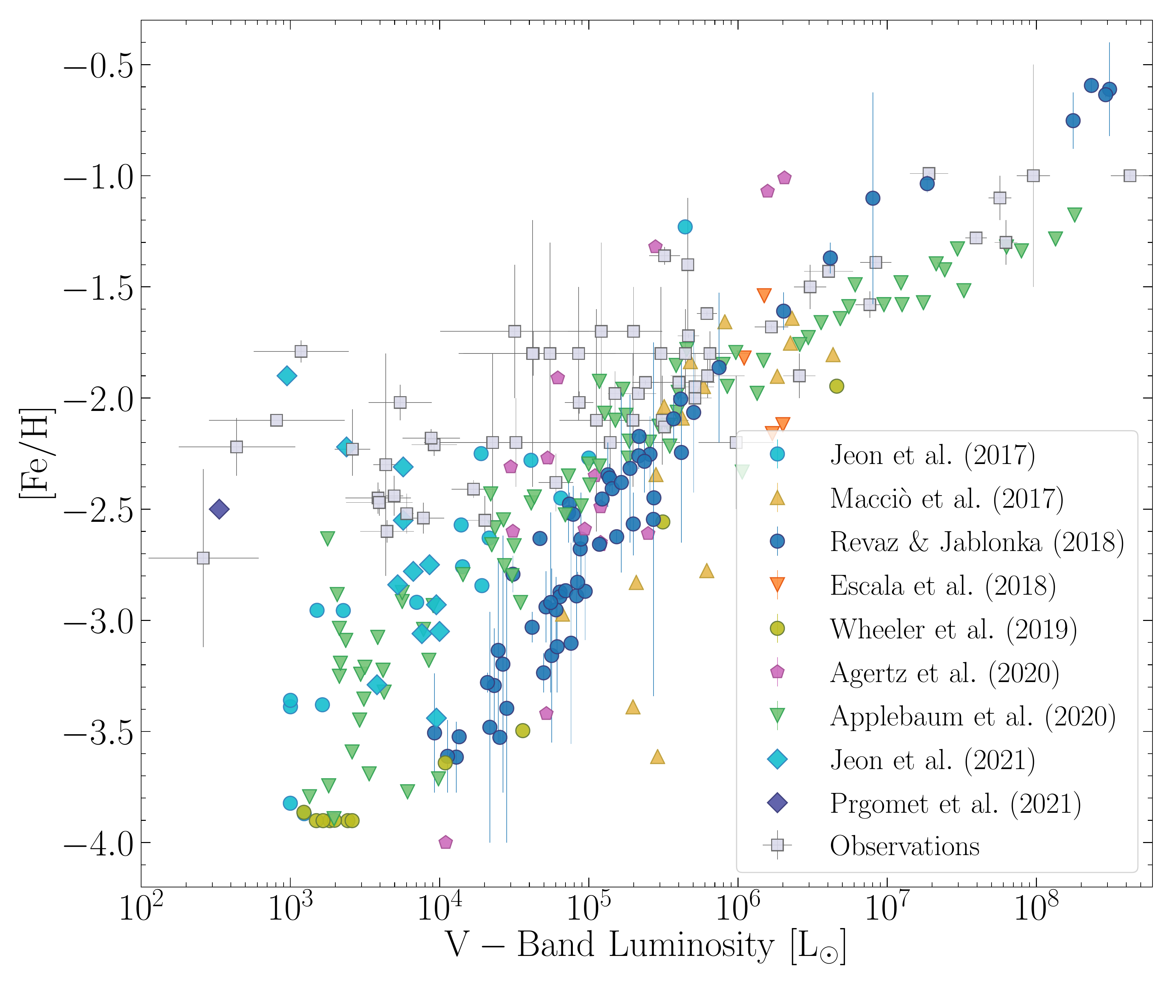}
  \caption{\small Comparison of the luminosity-metallicity relation for dwarfs and UFDs, between Local Group observations and simulations. The grey squares represent our Local Group sample (see text for details). 
  Coloured points stand for different simulations published with the references given in the bottom of the figure. We extended our own work \citep{Revaz2018} with the
  results of unpublished galaxies in the UFDs regime.
  Error bars for the metallicity in our simulated galaxies are calculated by comparing the results 
  obtained with the mean, the median and the mode acquired with several binning choices. See Section~\ref{sec:obs} for more
  details on the method used.
  \label{fig:ZL_base}
  }
  
\end{figure}

This discrepancy is illustrated in Fig.~\ref{fig:ZL_base}, where Local Group 
dwarfs and UFDs are compared 
to the available hydro-simulations from \cite{jeon2017,Maccio2017,Revaz2018,Escala2018,Wheeler_2019,agertz2020,applebaum2020,prgomet2022}.
This figure  includes a set of UFDs not published in \citet{Revaz2018} but obtained with exactly the same physical model.
The observed galaxy sample is obtained thanks to the continuously updated
\emph{Local Group and Nearby Dwarf Galaxies} data base
\footnote{\url{http://www.astro.uvic.ca/~alan/Nearby_Dwarf_Database.html}}
of \citet{McConnachie_2012} and are represented by grey squares.  
Note that our sample includes only confirmed UFDs that benefit from medium resolution spectroscopy with metallicity 
derived either from spectral synthesis or Calcium triplet (CaT) calibration.

%


Several explanations at the origin of this tension are proposed.
\citet{bovill2011a} suggest that the observed dwarfs may be a subset of bright primordial fossils which have 
been stripped of 90\% - 99\% of their stars. Indeed, tidal striping could 
remove stars from a dwarf, subsequently reducing its luminosity without affecting 
its metallicity.
However, only dwarfs with very small pericenter will see an appreciable fraction 
of their stars to be lost \citep{penarrubia2008}. 
Moreover, a striping would induce an expansion of the stellar component, in
contradition with UFDs having a very small half-light radius.
Environmental pre-enrichment by the host galaxy is suggested by \citet{Wheeler_2019}.
\citet{applebaum2020} mention however that in a Milky Way like environment, dwarf galaxies that have never been harassed or stripped show comparable metallicities with respect to the  dwarf satellites. 
They suggest instead that their simulated dwarfs are simply under-producing iron and that the tension can be solved by increasing the contribution to the iron abundance of Type Ia supernovae (SNeIa). This however requires to either reduce the SNeIa  explosion time scale, or extend the star formation history of UFDs over more than two $\mathrm{Gyr}$, in contradiction with current observations.


\citet{prgomet2022} look into the impact of a metallicity-dependent IMF on the stellar mass and metallicity of their simulated dwarf galaxies.  They examine the effect of a top-heavy IMF for metal-poor stars. 
They show that the largest number of massive stars, exploding as core collapse supernovae (CCSNe), decreases the final luminosity of the system through enhanced stellar feedback, while it increases the metallicity. Unfortunately, 
the boosted SNe feedback also leads to  highly diffuse systems with half-light radius $\sim0.5$ dex larger than the observed values.

Alternatively, \citet{Revaz2018} suggest that Pop III stars could play a role in reconciling observations with simulations,
a feature worth exploring in hydro-codes. Indeed, hydro-codes self-consistently treat the formation of galaxies, in particular
the merger history of sub-halos that may have been enriched independently. Moreover, they
properly account for the ejecta of polluted gas and its recycling after re-accretion on the same 
halo it originates or on different ones. 

\section{Methods and simulations}\label{sec:simulations}

%

We perform a large set of 19 zoom-in $\Lambda$CDM cosmological simulations of dwarf and UFD galaxies with a final
luminosity between $2\cdot 10^3$ to $3\cdot 10^5\,\rm{L_\sun}$.
We select galaxy halos from the $(3.4\,{\rm Mpc}/h)^3$ cosmological volume used in \citet{Revaz2018}.
We re-simulate those halos using a zoom-in technique including a full treatment of baryons that we will
describe below.
The initial conditions are obtained using the \texttt{MUSIC} code \citep{Hahn_2011}. 
Compared to our previous studies of dwarf galaxies \citep[e.g.][]{Revaz2018,harvey2018,hausammann2019,sanati2020}, simulations in this work are run 
with a factor of 8 enhancement in resolution for the gas and dark matter and a factor of 16 for the stars. This is essential to resolve the Pop III host halos with masses between $10^6$ and $10^7\,\mathrm{M_{\odot}}$ and to guarantee that UFDs with a luminosity as low as a few $10^3\,\rm{L_{\odot}}$ are populated with at least ten stellar particles.
In the refined region, we achieve a dark matter, gas and stellar mass resolution of 
$4172\,{\rm M_{\odot}}$,
$761\,{\rm M_{\odot}}$ and
$380\,{\rm M_{\odot}}$ respectively. 
Note that we limit our stellar resolution to $380\,{\rm M_{\odot}}$ to 
assure a proper IMF sampling up to the most massive stars in the mass range $[140-300]\,\mathrm{M_{\odot}}$ considered in this work (see Sec.~\ref{sec:yields} and \ref{sec:IMF}). 

We use the cosmology of the  \cite{Planck2016} with $\Omega_\Lambda = 0.685$,  
$\Omega_m = 0.315$, $\Omega_b = 0.0486$, $H_0 = 67.3\,\textrm{km s}^{-1} \textrm{Mpc}^{-1}$, $n_s = 0.9603$
and $\sigma_8 = 0.829$. All simulations are started at redshift $z = 70$, ensuring that the rms variance of the initial density field, $\sigma_{8}$, lies between 0.1 and 0.2 \citep{knebe2009,onorbe2014}, and run till redshift $z = 0$.


\subsection{Numerical setup}\label{sec:numerical setup}

The simulations are run with \textsc{GEAR} \citep{Revaz2012, Revaz2016,Revaz2018}, 
a fully parallel chemo-dynamical Tree/Smoothed Particle Hydrodynamics(\texttt{SPH}) code based on \textsc{Gadget-2} \citep{Springel_2005}.
In addition to the collisionless dark matter, \textsc{GEAR} treats the baryonic physics by including gas cooling,  redshift-dependent UV-background heating, star formation and stellar feedback. 
We briefly summarize its essential elements below.
  
\paragraph{Radiative cooling} 
Gas cooling is computed using the \textsc{GRACKLE} library \citep{Smith2017}. In addition
to primordial gas cooling, it includes metal-lines cooling scaled according to the gas metallicity.
\textsc{GRACKLE} also includes UV-background radiation based on the prediction from \citet{haardt2012}. Hydrogen self-shielding against the ionizing radiation is incorporated by suppressing the UV-background heating  for gas densities above $n_{\rm{H}} = 0.007$ $\mathrm{cm}^{-3}$ \citep{aubert2010}.  
 
\paragraph{Star formation} 
Star formation is modelled using the stochastic prescription proposed by \citet{katz1992} and \citet{katz1996} that reproduces the Schmidt law \citep{1959ApJ...129..243S}. This classical recipe is supplemented by a modified version of the Jeans pressure floor through adding a non-thermal term in the equation of state of the gas. 
The purpose of this modification is to avoid any spurious gas fragmentation \citep{truelove1997,bate1997,owen1997}.
%
%
In dense regions where the system is dominated by unresolved physics we adopt a star formation density threshold based on the Jeans polytrope, directly correlating the \texttt{SPH} smoothing length with density:
\begin{equation}
\rho_{\mathrm{SFR},i} = \frac{\pi}{4}G^{-1}{N_\mathrm{Jeans}}^{-2/3} {h_{i}}^{-2} \left( \gamma \frac{K_B}{\mu m_{\mathrm{H}}} T + {\sigma_i}^2\right).
\label{eq:pressure}
\end{equation}
Here $G$ is the universal gravitational constant and $\gamma$ the addiabatic index of the gas fixed to $5/3$. $h_{i}, \rho_{i}$ are respectively the \texttt{SPH} smoothing length and density.
$\sigma_{i}$ is the velocity dispersion of gas particle $i$, summed over the neighboring particles in the Jeans mass.
$N_{\mathrm{Jeans}}$ is the ratio between the Jeans mass and the \texttt{SPH} mass resolution of particle $i$ and
is fixed to 10.
In this work, we supplement this temperature dependent threshold with
a constant density threshold $\rho_{\rm SFR,c}$ which prevents stars to form in the cold low-density gas regions. The importance of this threshold and 
its value will be discussed in Sec.~\ref{sec:cal}.
Above $\max\,(\rho_{\rm SFR,c},\rho_{\rm SFR,i})$ stars may form with a star formation efficiency of $1\%$. 

Depending on the metallicity of the gas particle from which it will form, the stellar particle is composed of either Pop II or Pop III stars.
Pop III stars only form from gas particles with $[\rm{Fe/H}]$ lower than a critical value $[\rm{Fe/H}]_{\rm c}$.  Above this metallicity only Pop II stars will form. 
As stars with $\rm[Fe/H]<-5$ are extremely rare in the observational data, for our fiducial model we set $[\rm{Fe/H}]_{\rm c}$ equal to $-5$. 
Following this choice, all stars more metal-poor than $\rm[Fe/H]=-5$ explode in a few $\rm{Myrs}$, and will be absent at redshift $z=0$.  
In Sec.~\ref{sec:MetallicityThreshold}, we discuss the impact of modifying this parameter.  
Because $[\rm{Fe/H}]_{\rm c}$ is not reached in all gas particles at the time, these Pop II and Pop III stars can potentially coexist at a given time step.


\subsection{Initial mass function (IMF)}\label{sec:IMF}



As Pop III stars are predicted to have their specific IMF and yields \citep[e.g.,][]{Iwamoto_2005,heger2010,Bromm_2013}, we thus updated \textsc{GEAR} with the possibility to fully distinguish the properties of Pop II and Pop III stars. 
In our approach, each stellar particle represents a single stellar population (SSP) that is associated with its own IMF
and stellar yields. 

For both Pop II and Pop III, the IMF is described by the probability function $\Phi(m)$, where $\Phi(m)\,dm$ gives the fraction of stars found in the mass range $[m,m+dm]$. In our implementation, $\Phi(m)$ is defined by a
set of power laws with the slope $\alpha(m)$ depending on the mass interval~:
\begin{equation}
  \Phi(m) = \frac{m^{{\alpha(m)}}}{\beta},
\end{equation} 
with $\beta$ the normalisation constant that ensures the integral of $\Phi(m)$ over the entire mass range considered to be 1.

\paragraph{Pop II:} 
We used the revised IMF of \cite{Kroupa_2001} (systematic bias due to unresolved binaries in observations is considered), previously implemented in \textsc{GEAR} with
     \begin{equation}
        \alpha(m)=\left\{ 
        \begin{array}{rcr}
         0.7, &\rm{if}& m \in [0.05\,\rm{M_\odot} ,0.08\,\rm{M_\odot}] \\
        -0.8, &\rm{if}& m \in [0.08\,\rm{M_\odot} ,0.50\,\rm{M_\odot}] \\
        -1.7, &\rm{if}& m \in [0.50\,\rm{M_\odot} ,1.00\,\rm{M_\odot}] \\
        -1.3, &\rm{if}& m \in [1.00\,\rm{M_\odot} ,50.0\,\rm{M_\odot}] \\
        \end{array}
        \right.
        \label{kroupa}
        \end{equation}
The minimal and maximal mass of stars are set to respectively $0.05$ and $50\,\rm{M_\odot}$. 

\paragraph{Pop III:}

Adopting a different IMF for Pop III than that of Pop II stars is motivated by the fact that no metal-free star has been detected so far, an indication that there is no or very few long-lived primordial low mass stars. It is thus expected that the majority of Pop III stars 
are massive enough to explode as CCSNe and disappear \citep{Larson1998,bromm1999,karlsson2008}. 
For Pop III stars, we adopt a simple power law IMF with a unique slope $\alpha=-1.3$, similar to the one of 
the most massive Pop II stars. The minimal mass $M_{\rm{min}}$ is set to $13\,\rm{M_\odot}$. It is worth mentioning that $M_{\rm{min}}$ is above the lower supernova mass limit.
The maximal mass $M_{\rm{max}}$, when PISNe are not considered, is set to $140\,\rm{M_\odot}$. When PISNe are included, $M_{\rm{max}}$
is set to $300\,\rm{M_\odot}$. 


\subsection{Stellar feedback: yields and energy}\label{sec:yields}


At each time step the exploding supernovae are randomly selected  
following the random discrete IMF sampling (RIMFS) scheme of \cite{Revaz2016}, in which the IMF is considered as a probability distribution. The stellar lifetimes are mass- and metallicity-dependent following \citep[][private communication]{kodama1997}.


%
\begin{figure}[h]
  \centering
  \includegraphics[width=0.49\textwidth]{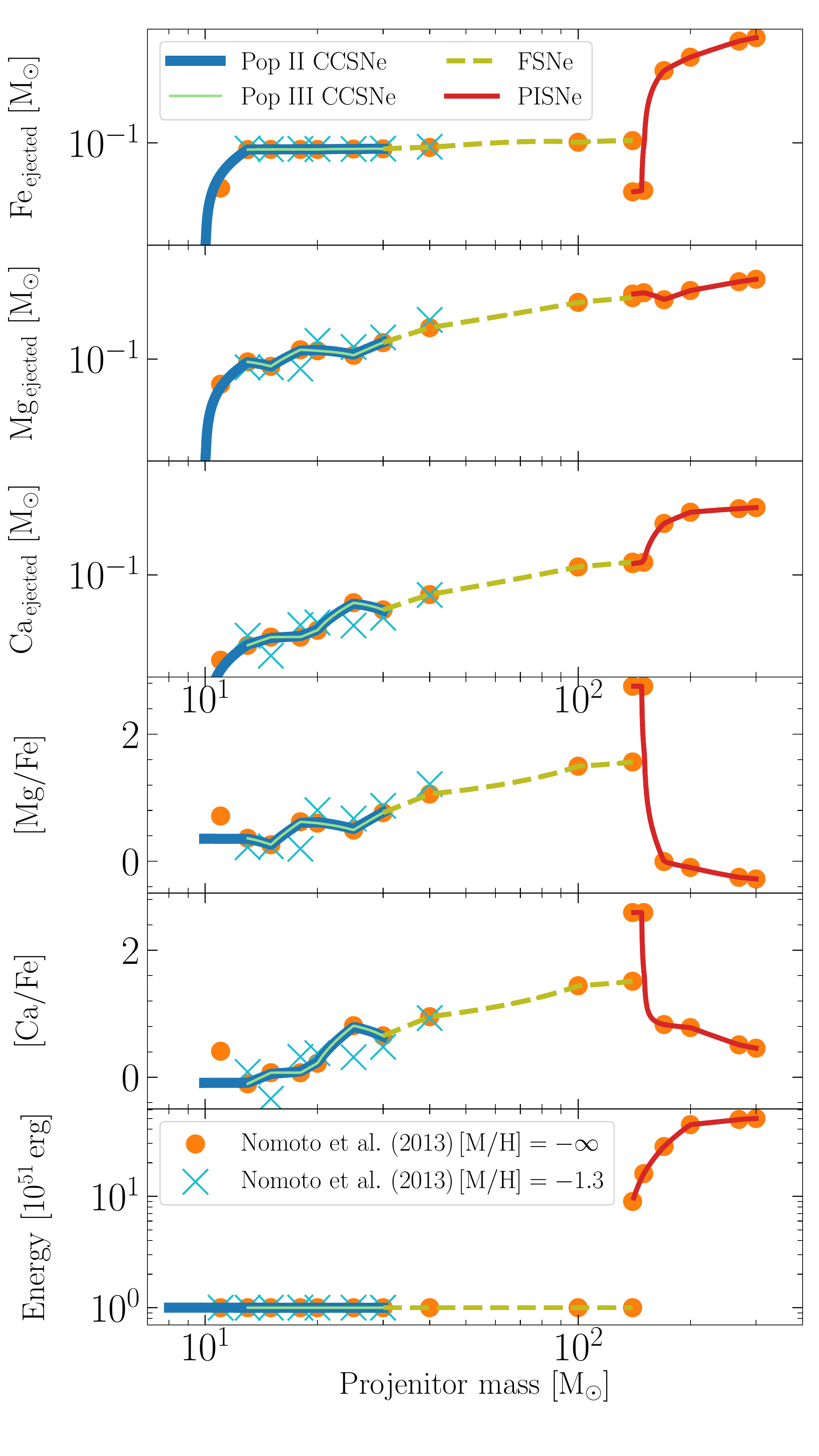}
  \caption{\small Comparison of the yields of iron, magnesium and calcium  as a function of  stellar masses  from \citetalias{NK13}. The corresponding abundance ratios [Mg/Fe] and [Ca/Fe] and the supernovae explosion energies are also shown. The \citetalias{NK13} table values for stars with $M_{\rm{proj}}<30\,\rm{M_{\odot}}$ at  
  $[\rm{M/H}]\,=-\infty$ and $-1.3$ are shown with orange circles and  blue crosses, respectively. The thick blue and thin green curves show the adopted, interpolated, values used in this work for, respectively, the Pop II and Pop III stars. 
  The dashed yellow lines indicates the mass range of stars ending as FSNe.
  The red curves show the yields and energy for PISNe with $M_{\rm{proj}}>140\,\rm{M}_\odot$ examined in this work 
  from \citetalias{NK13}.}
\label{fig:fe_sum}
\end{figure}
%
As to the massive stars, our previous works \citep{Revaz2012,Nichols2014,Nichols2015, Revaz2016,Revaz2018,harvey2018,hausammann2019,sanati2020}, 
only included  CCSNe, with the yields of  \citet{tsu95}, for stellar masses up to  $70\,\rm{M_{\odot}}$. We had not yet considered zero metallicity stars.
Therefore, for the purpose of the present study, for which we do need Pop III stars, we introduce the metallicity-dependent yields of \citetalias{NK13}, from $\rm{[M/H]}=-\infty$ to $0.4$.
Figure~\ref{fig:fe_sum} summarizes the yields and feedback energy of Pop II and Pop III stars, as considered in this study and further described in below. 

\paragraph{Pop II:} 
As indicated above, we use the yield tables of \citetalias{NK13} taking into account the analysis and conclusions of \citet{kobayashi2020}. 
The \citetalias{NK13} yields are provided in bins of metallicty at $\rm{[M/H]}=-\infty, -1.3,-0.7,-0.4,0,0.4$.
As \textsc{GEAR} only treats metallicity-independent yields, we interpolate the \citetalias{NK13} tables at  $\rm{[M/H]}=-2.5$, which globally corresponds to the average metallicity of UFDs. The result of this interpolation is seen with thick blue lines in Fig.~\ref{fig:fe_sum}.
Below $40\,\rm{M_\odot}$ the values of the \citetalias{NK13} table for the two lowest metallicities, $\rm{[M/H]}=-\infty$ and $\rm{[M/H]}=-1.3$ are shown with orange circles and blue crosses, respectively. The proximity of these two set of values (barely visible for the iron ejecta) supports our interpolation.

As in \citet{kobayashi2020} 
only supernovae progenitors less massive than $30\,\rm{M_\odot}$ will contribute to the chemical enrichment.
Above this mass, numerical models predict that the formation of supernova progenitors is quite unlikely \citep{2009ARA&A..47...63S}.
We thus ignore the explosion of those stars which are referred to as failed supernovae (FSNe) in the text.
For all Pop II CCSNe with mass less than $30\,\rm{M_\odot}$ the energy released per explosion is fixed to the fiducial value of $10^{51}\,\mathrm{erg}$. 

\paragraph{SNeIa:} 
Stars with initial masses between $3$ and $8\,\rm{M_{\odot}}$ can be progenitors of Type Ia supernovae (SNeIa), following the model of binary systems of \citet{kobayashi2000}.  The SNeIa yields are taken from \citet{tsu95}. Their explosion energy is fixed to $10^{51}\,\rm{erg}$.

\paragraph{Pop III:} The yields and energy of Pop III stars are those of the metal-free 
($\rm{[M/H]}=-\infty$) stars provided by the \citetalias{NK13}. 
In mass range $13$ to $30\,\rm{M_\odot}$, shown with thin green lines in Fig.~\ref{fig:fe_sum}, Pop III stars explode as CCSNe with feedback energy $10^{51}\,\mathrm{erg}$.
As the uncertainties in the physics of black hole formation from the first massive stars obscure the upper limit of type II supernovae, for Pop III with initial masses between $30$ and $140\,\rm{M_{\odot}}$ (dashed yellow lines, 
we examine two different final fates:  
(i) normal CCSNe with nucleosynthesis yields from \citetalias{NK13}, 
(ii) FSNe which according to \citet{kobayashi2020} end as black holes. In this latter case, no energy nor yields are ejected.




In the PISNe regime for stars above $140\,\rm{M_\odot}$, yields are also taken from the \citetalias{NK13} table, corresponding to the model predictions of \citet{umeda2002}.
In this mass regime, shown in red lines, supernova energy varies
from $9$ to $50\,\cdot\,10^{51}\,\mathrm{erg}$ for the most massive stars with $300\,\mathrm{M}_\odot$.


%


%
\begin{table*}[h]
  \caption{\small
  Parameters of the simulations. 
  Columns are as follows: 
  1) Model description and ID.  
  2) Mass range of Pop II and Pop III stars. 
  3) IMF slopes. 
  4) Mass range of Pop II and Pop III stars fall into a CCSNe, for which a feedback energy of $10^{51}\,\mathrm{erg}$ is assumed. 
  5) Mass range of Pop II and Pop III stars which fail to explode and fall into black holes. 
  6) Mass range of Pop III stars forming PISNe and the feedback energy released after their explosion.
  7) Critical metallicity below which only Pop III stars form. Above this metallicity only Pop II stars form. 
  8) Density threshold in gas clouds $\rho_{\rm SFR,c}$ required for the onset of star formation.
  9) Radius $R_{\mathrm{ejecta}}$ around each supernova in which metals are ejected.
  \label{tab:params}}
  \centering
  \resizebox{0.95\textwidth}{!}{
  \begin{tabular}{l c c c c c c c c }
  \hline
  \hline
     (1) & (2) & (3) & (4) & (5) & (6) & (7) & (8) & (9) \\
  Model description & Mass range $[\mathrm{M_{\odot}}]$  & IMF slope($\alpha$) & CCSNe  & FSNe & PISNe & Critical & Density & Metal ejection \\
  
  \texttt{Model ID} & Pop II  & Pop II & Pop II  & Pop II & $\mathrm{M}\,[\mathrm{M}_{\odot}]$ &  metallicity & threshold & radius \\
  
           & Pop III  & Pop III  & Pop III &    Pop III  &  $\mathrm{E}\,[10^{51}\mathrm{erg}]$    &  $[\rm{Fe/H}]_{\rm{c}}$ & $\rho_{\mathrm{SFR},c}\,[\rm{atom}/\mathrm{cm^3}]$ & $R_{\mathrm{ejecta}}[\mathrm{pc}]$  \\
  \hline
  \hline
  \multicolumn{6}{l}{Models with only Pop II stars} \\
 \texttt{PopII} & $[0.05, 50]$ & $[0.7,-0.8,-1.7,-1.3]$ & $[8, 30]$ &   $[30, 50]$ & - &  - & 5 & $300$ \\
    & - & -  & - & -  & - & &  &  \\
  \hline
  \multicolumn{6}{l}{Models with Pop III stars} \\
    \texttt{PopII+PopIII(CCSNe)}  & $[0.05, 50]$ & $[0.7,-0.8,-1.7,-1.3]$ & $[13, 30]$ &   $[30, 50]$  & - & $-5$ & 5 & $300$ \\
    &   $[13, 140]$ &  $-1.3$ & $[13, 140]$  &    -  & - & & & \\
  \hline
  \multicolumn{6}{l}{Models with Failed SNe} \\
\texttt{PopII+PopIII(CCSNe+FSNe)}   &  $[0.05, 50]$ & $[0.7,-0.8,-1.7,-1.3]$ & $[13, 30]$ &   $[30, 50]$  & - & $-5$ &  1 , 5 & $h_{i}$ , $300$ \\
     & $[13, 140]$  &  $-1.3$    & $[13, 30]$ &   $[30, 140]$  & - &  & &  \\
  \hline
  \multicolumn{6}{l}{Models with PISNe} \\
   \texttt{PopII+PopIII(CCSNe+FSNe)+PISNe}  & $[0.05, 50]$ & $[0.7,-0.8,-1.7,-1.3]$ & $[13, 30]$ &   $[30, 50]$  & $[140, 300]$ & $-5$ & 5 & $300$ \\
   &  $[13, 300]$  & $-1.3$  & $[13, 30]$ &   $[30, 140]$  & $[9, 50]$ &  & &    \\
  \hline
  \multicolumn{6}{l}{Models with decreasing PISNe feedback energy} \\
 \texttt{PopII+PopIII(CCSNe+FSNe)+PISNe($\texttt{E}_{\texttt{51}}$)} & $[0.05, 50]$ & $[0.7,-0.8,-1.7,-1.3]$ & $[13, 30]$ &   $[30, 50]$  & $[140, 300]$ & $-5$ & 5 & $300$ \\
   &  $[13, 300]$  & $-1.3$  & $[13, 30]$ &   $[30, 140]$  & $1$ &  & &   \\
  \hline
  \multicolumn{6}{l}{Models with increasing critical metallicty} \\
   \texttt{PopII+PopIII(CCSNe+FSNe)+PISNe(Z-4)} & $[0.05, 50]$ & $[0.7,-0.8,-1.7,-1.3]$ & $[13, 30]$ &   $[30, 50]$  & $[140, 300]$ & $-4$ & 5 & $300$ \\
   &  $[13, 300]$  & $-1.3$  & $[13, 30]$ &   $[30, 140]$  & $[9, 50]$  &  & &    \\
  \hline
  \hline
  \end{tabular}}

\end{table*}
\subsection{Stellar feedback}\label{sec:ejection}

Exploding supernovae release both synthesized chemical elements and energy into the ISM. In particle-based codes, this is modeled by distributing these quantities
on the neighboring particles, whose definition and identification is not a trivial. We describe here the two methods implemented in \textsc{GEAR} and discuss their 
advantages and drawbacks. The impact of these schemes as well as their parameters will
be studied in Sec.~\ref{sec:cal}.

\paragraph{Injecting into the \texttt{SPH} kernel}

The standard method consists in releasing the ejecta into the effective number of neighboring particles $N_{\rm{eff}}$ which is determined by the implemented \texttt{SPH} scheme.
In the modern conservative formulation of \texttt{SPH} \citep{springel2002,price2012,springel2010,hopkins2013} as 
used by \textsc{GEAR}, a differenciable 
relation between the volume of a particle (or equivalently its density) 
and its smoothing length is required. The most natural one is to consider that mass in the kernel volume of a 
particle remains constant, equal to $\bar{N}_{\rm{ngb}}$ times the mass $m_i$  of particle $i$.
The parameter $\bar{N}_{\rm{ngb}}$ (fixed to 50) sets the average number of 
neighboring particles defined by:
\begin{equation}
\frac{4}{3}\,\pi h^3_i \rho_i = m_i \bar{N}_{\rm{ngb}},
\end{equation}
where the smoothing length $h_i$ and density $\rho_i$ of particle $i$ are related to the volume $V_i$  by
$\rho_i = m_i/V_i$.
Defining $V_i$ through the summation over neighboring particles, via the kernel function $W$ \footnote{Note that there are different possibilities to express $V_i$, see \citet{hopkins2013}.}:
\begin{equation}
V_i =  \frac{1}{\sum_{j=1}^{N_{\rm{eff}}} W( |\vec{r}_i-\vec{r}_j |,h_i)},
\label{eq:Nngb}
\end{equation}
one can explicitly express $\bar{N}_{\rm{ngb}}$:
\begin{equation}
\bar{N}_{\rm{ngb}} = \frac{4}{3}\,\pi h^3_i \sum_{j=1}^{N_{\rm{eff}}} W( |\vec{r}_i-\vec{r}_j |,h_i).
\label{eq:Nngb}
\end{equation}
This latter relation shows that the effective number of neighboring particles $N_{\rm{eff}}$ is different than the fixed parameter $\bar{N}_{\rm{ngb}}$, and that $N_{\rm{eff}}$ strongly depends on the distribution of the neighboring particles through $W( |\vec{r}_i-\vec{r}_j |,h_i)$.
We thus expect this number to be different for each time ejection event. 

A particular case worth mentioning is when two supernovae explode subsequently in the same region. In this case, the second supernova explode in the Sedov cavity created by the first explosion. Assuming an initial homogeneous ISM, for the first explosion, the \texttt{SPH} scheme will find 
$N_{\rm{eff}} \cong \bar{N}_{\rm{ngb}}$, while for the second supernova, $N_{\rm{eff}}$ will largely  exceeds $\bar{N}_{\rm{ngb}}$. 
Indeed, as neighboring particles can only be found at larger distances, each one of them will have a low $W( |\vec{r}_i-\vec{r}_j |,h_i)$. In this case, a large number of particles have to be considered in the sum of Eq.~\ref{eq:Nngb}, in order to guarantee the convergence towards the $\bar{N}_{\rm{ngb}}$.
Consequently, this scheme can potentially causes the ejection of metals into unrealistic and uncontrolled distances.
Moreover, it is strongly resolution dependent as $h_i$ scales with the spatial resolution. 
We further discuss the results of using this scheme in Sec.~\ref{sec:cal}. 

\paragraph{Injecting into a constant volume}

An alternative to the previous scheme is to release the ejecta among particles that reside in a constant volume. This volume is defined by a sphere of radius $R_{\mathrm{ejecta}}$ around the exploding star.
In addition to avoiding the bias of ejecting into the \texttt{SPH} volume, the method is motivated
by reproducing the effective extent of a supernova explosion.
The adiabatic supernova explosion in a homogeneous medium must follow the Sedov-Taylor solution 
\citep{1959sdmm.book.....S,1950RSPSA.201..159T}. The mean pressure $\overline{P}$ within the blast wave falls 
off rapidly with radius. This means that the expansion of the supernova shock must follow a power law which 
at the end of the blast wave phase, 
using typical values for supernova explosions, 
is approximately equal to:
\begin{eqnarray}
R_{\mathrm{ejecta}} \approx
     300\, \mathrm{pc}\, {\left( \frac{E}{10^{51}\,\mathrm{erg}} \right)}^{1/3} {\left(\frac{\overline{P}} {4\times10^{-13} \, \mathrm{dyne} \, \mathrm{cm}^{-2}} \right) }^{-1/3},     
\end{eqnarray}
%
where we assume that an amount of energy $E$ is instantaneously ejected into the ambient medium of 
uniform density $\rho$, characterized by 
the adiabatic index $\gamma=\frac{5}{3}$ \citep{1988RvMP...60....1O}. 
We thus supplemented \textsc{GEAR} with the possibility to eject metals in a constant physical volume
(not a comoving volume) 
using $R_{\mathrm{ejecta}}$ that we choose to be $300\,\rm{pc}$. 


\paragraph{Delayed cooling and mixing}
In both schemes mentioned above, deposited energy and metals are weighted by
the kernel $W( |\vec{r}_i-\vec{r}_j |,h_i)$.
To avoid an instantaneous radiation of the injected energy,
we use the delayed cooling method which consists in 
disabling gas cooling for a short period of time \citep{stinson2006}, here taken as $5\,\rm{Myr}$.
Only $\epsilon$ fraction of the supernova energy 
is deposited into the ISM. We assume $\epsilon= 0.2$, hence $\sim$20\% of the released energy impacts the ISM effectively. 
The released chemical elements are further mixed in the ISM using the Smooth metallicity scheme 
\citep{okamoto2005,tornatore2007,wiersma2009}. Basically, the chemical abundance of a gas particle results  from the convolution of the metal content of its neighbors.

\subsection{Models}\label{sec:models}

We perform a set of simulations in order to explore the impact of the different parameters listed in Tab.~\ref{tab:params}.
The stellar range of properties as well as the model ID are provided  in the first column.

Models including only Pop II stars set our reference, hereafter \texttt{PopII}. 
The  mass range of Pop II and Pop III stars is given in the second column. 
As described in Sec.~\ref{sec:IMF}, the IMF slope of the Pop II stars changes with the stellar mass range \citep{Kroupa_2001}. For the Pop III stars, we adopt a fixed slope of $-1.3$ similarly to  the most massive of the Pop II stars. 

The mass range of CCSNe is shown in the forth column.
In \texttt{PopIII(CCSNe)}, all Pop III stars explode as a normal CCSNe, while in \texttt{PopIII(CCSNe+FSNe)}, stars in mass range $[30-140]\,\mathrm{M}_{\odot}$ fail to explode and fall into black holes. 
The mass range of these FSNe (see Sec.~\ref{sec:yields}), ejecting no energy nor yields, is given in the fifth column. 

For massive Pop III stars ($[140-300]\,\mathrm{M}_{\odot}$), exploding as PISNe, we explore the two feedback energy prescriptions indicated in the sixth column: (i) In our reference \texttt{PISNe} model, the highly energetic feedback is increasing from $9$ to $50\cdot10^{51}\,\mathrm{erg}$, according to the initial mass of the star \citepalias[see][for more details]{NK13}.
(ii) In \texttt{PISNe($\texttt{E}_{\texttt{51}}$)} model, we explore a less energetic feedback. 
In this model, the supernova energy is independent of the progenitor mass and equal to the energy ejected by the normal CCSNe, i.e.,  ($10^{51}\,\mathrm{erg}$). 

The critical metallicity $[\mathrm{Fe/H}]_{c}$, below which only the first massive and metal-free stars may form, is shown in the seventh column. When the metallicity of gas is higher than this critical value, low-mass long-lived Pop II stars can form. In model \texttt{PISNe(Z-4)}, we increase $[\mathrm{Fe/H}]_{c}$ from $-5$, of   our reference \texttt{PISNe} model, to $-4$.

In model \texttt{PopIII(CCSNe+FSNe)}, we explore the impact of the density threshold in gas clouds, $\rho_{\rm SFR,c}$, required for the onset star formation. In this model we also explore the radius $R_{\mathrm{ejecta}}$ around each supernova in which metals are ejected (see Sec.~\ref{sec:ejection}). These two parameters are given in columns eight and nine. In all other models, these parameters are fixed to $5\,{\rm atom}/\mathrm{cm}^3$ and $300\,\rm{pc}$, respectively. 


\subsection{Extraction of the Observables, luminosity and metallicity}\label{sec:obs}

Both the galaxy luminosity and metallicity are extracted from the low mass long-lived stars traced by the Pop II stellar particles. 
All Pop III stars explode and therefore do not contribute to the final stellar mass budget, luminosity or metallicity of their host galaxy. 
The line-of-sight (LOS) stellar velocity dispersion, $\sigma_{LOS}$, is calculated for seven different lines of sight inside a $1\,\rm{kpc}$ cylindrical radius. The value quoted for each galaxy represents the mean of these values.

The galaxy V-band luminosity ($L_V$) is obtained by considering all Pop II stellar particles falling within the $90\%$ of the halo light-radius. Their mass is converted into luminosity using the stellar population synthesis model of \citet{vazdekis1996}  computed with the revised \citet{Kroupa_2001} IMF. Where necessary, the luminosities are inter- and extra-polated in age and metallicity using a bivariate spline.

Because of the limited number  of stars formed in UFDs,
the choice of a representative \textit{mean} $[\rm{Fe/H}]$ is not a trivial task. As the metallicity distribution is sparsely sampled computing the mode, the peak of the metallicity  distribution function can lead to large uncertainties. 
In Appendix~\ref{sec:appendix1}, we propose a method to determine the mode based  on a fitting analytical formula derived from a simple chemical evolution model. The error on  $[\rm{Fe/H}]$ is taken at the maximum of the errors obtained by the different methods, peak and mode. All chemical abundances are calculated with respect to the solar abundances of \citet{anders1989}.

%

\section{Results}\label{sec:results}

Table~\ref{tab:catalogue} summarizes the main properties of the simulated halos  at $z=0$ for  our fiducial model which includes only Pop II stars. The halo IDs are those of \citet[][see their Tab.~1]{Revaz2018}, while the simulations are run with the physics described in this paper. Table~\ref{tab:catalogue} provides the galaxy total V-band luminosity $L_{\mathrm{V}}$,   total stellar mas $M_{\star}$,  virial mass $M_{200}$,  virial radius 
$R_{200}$,  mean stellar velocity dispersion $\sigma_{\rm LOS}$ and  peak value of the metallicity distribution function $[\rm{Fe/H}]$.
The halo \texttt{h177} 
hosts a Sextans-like classical dwarf galaxy (see next Section).
The other 18 halos host dwarfs with luminosities equal or lower than $10^5\,\rm{L_{\odot}}$ and are considered as UFDs.

\begin{table}[h]
  \caption{\small Global properties of the 19 halos simulated when only Pop II stars are considered. The first column gives the halo ID following \citet{Revaz2018}. $L_{\rm{V}}$ is the V-band luminosity and $M_{\star}$ the stellar mass. $M_{\rm{200}}$ is the virial mass, i.e., the mass inside the virial radius $R_{\rm{200}}$. $r_{1/2}$ is the half-light radius, $\sigma_{\rm{LOS}}$ is the line-of-sight velocity dispersion and $[\rm{Fe/H}]$ is the peak value of the metallicity distribution function. 
    \label{tab:catalogue}}
  \centering
  \resizebox{0.49\textwidth}{!}{%
  \begin{tabular}{l c c c c c c c }
  \hline
  \hline

    Halo ID & $L_{\rm{V}}$           & $M_{\star}$            & $M_{\rm{200}}$          & $R_{\rm{200}}$  &  $r_{1/2}$ & $\sigma_{\rm{LOS}}$ & $[\rm{Fe/H}]$ \\
       & $[10^5\,\rm{L_\odot}]$ & $[10^5\,\rm{L_\odot}]$ & $[10^9\,\rm{M_\odot}]$  & $[\rm{kpc}]$  & $[\rm{kpc}]$  & $[\rm{km/s}]$ & [dex] \\ 
  \hline
  \hline
\texttt{h177} & 3.07 & 10.28 & 0.51 & 25.9 & 0.26 & 5.7 & $-1.97$ \\
\texttt{h063} & 1.41 & 3.73 & 2.41 & 41.8 & 0.51 & 6.4 & $-2.31$ \\
\texttt{h190} & 1.31 & 3.62 & 0.47 & 24.2 & 0.28 & 4.8 & $-2.15$ \\
\texttt{h187} & 1.04 & 2.86 & 0.43 & 23.4 & 0.25 & 5.3 & $-2.27$ \\
\texttt{h166} & 0.71 & 1.92 & 0.64 & 26.9 & 0.22 & 5.0 & $-2.36$ \\
\texttt{h112} & 0.63 & 1.68 & 1.18 & 32.9 & 0.43 & 6.9 & $-2.79$ \\
\texttt{h152} & 0.53 & 1.42 & 0.80 & 28.9 & 0.38 & 6.4 & $-2.82$ \\
\texttt{h113} & 0.41 & 1.08 & 1.14 & 32.5 & 0.53 & 7.4 & $-3.07$ \\
\texttt{h184} & 0.30 & 0.81 & 0.61 & 26.4 & 0.28 & 4.8 & $-2.69$ \\
\texttt{h245} & 0.17 & 0.45 & 0.39 & 22.7 & 0.32 & 6.0 & $-3.24$ \\
\texttt{h259} & 0.20 & 0.53 & 0.39 & 22.7 & 0.21 & 5.1 & $-2.74$ \\
\texttt{h226} & 0.15 & 0.38 & 0.44 & 23.6 & 0.25 & 4.8 & $-2.87$ \\
\texttt{h277} & 0.10 & 0.26 & 0.36 & 22.2 & 0.42 & 5.3 & $-3.03$ \\
\texttt{h249} & 0.07 & 0.18 & 0.39 & 22.8 & 0.28 & 5.7 & $-3.18$ \\
\texttt{h315} & 0.05 & 0.14 & 0.33 & 21.6 & 0.51 & 5.6 & $-3.16$ \\
\texttt{h323} & 0.05 & 0.13 & 0.27 & 20.1 & 0.21 & 4.0 & $-3.07$ \\
\texttt{h170} & 0.05 & 0.13 & 0.62 & 26.5 & 0.41 & 5.0 & $-3.56$ \\
\texttt{h273} & 0.04 & 0.09 & 0.39 & 22.7 & 0.32 & 4.1 & $-3.53$ \\
\texttt{h291} & 0.02 & 0.04 & 0.33 & 21.4 & 0.18 & 5.5 & $-3.69$ \\
  \hline
  \hline
  \end{tabular}%
  }

\end{table}

\begin{figure*}[!h]
    \centering
    \includegraphics[width=\textwidth]{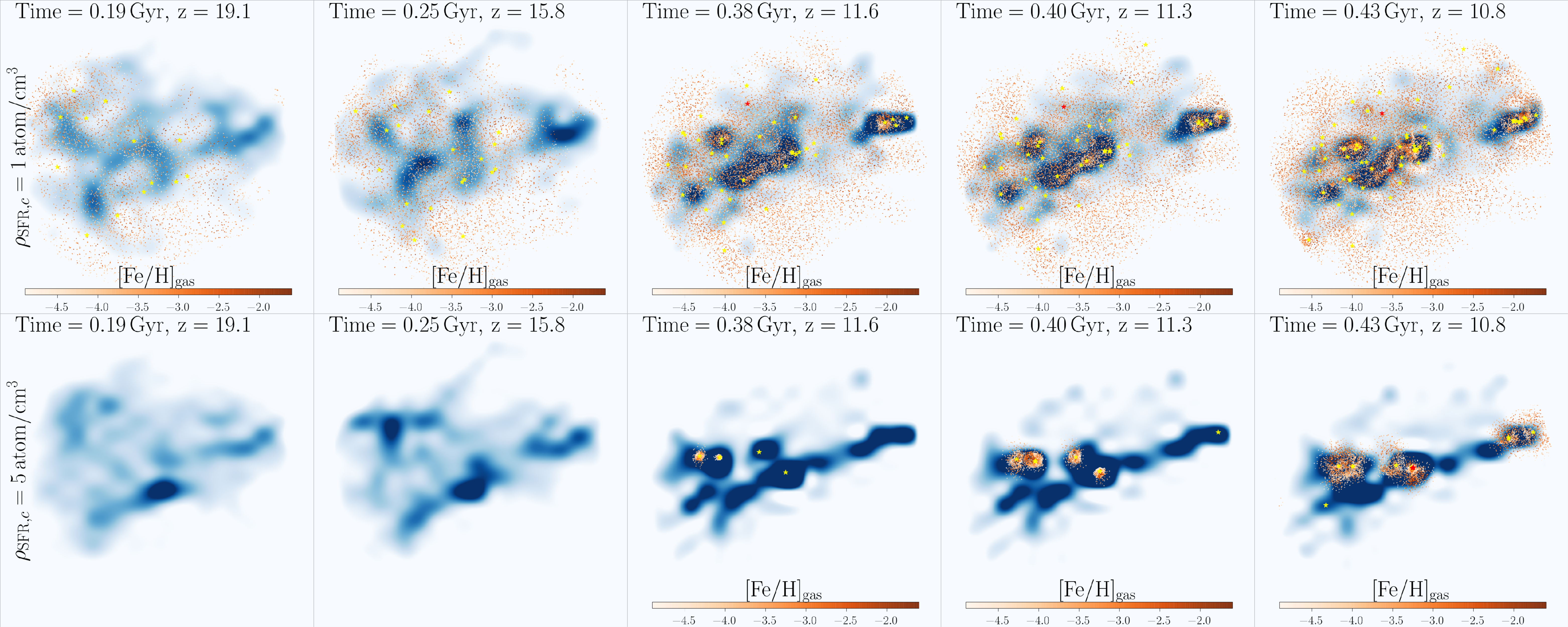}
  \caption{\small Metal enrichment of the Sextans-like model galaxy, \texttt{h177}, from $190$ to $430\,\mathrm{Myr}$. The surface density of gas is color-coded in blue. After each supernova explosion, shown in yellow and red for Pop III and Pop II stars, yields are ejected, using the \texttt{SPH} kernel scheme. 
  The metallicity of gas, color-coded in orange, is determined by the amount of Fe received from the supernovae ejecta. 
  The star formation density threshold $\rho_{\mathrm{SFR},c}$ is set to, respectively, $1$ and $5\,\rm{atom/cm^{3}}$ in the top and bottom panels.}
  \label{fig:TimeEvolv}
\end{figure*}
%


\subsection{Model calibration and validation}\label{sec:cal}

While in \citet{Revaz2018} we could successfully  reproduce the observed properties of the Local Group  classical dwarf galaxies, in the present study, not only 
we use different nucleosynthesis tables, but we also  increase the 
effective resolution of the simulations by one order of magnitude. Therefore, before taking up the challenge of UFDs, we first need to calibrate our physical 
reference model. To this end, we focus on the halo \texttt{h177}, representative of a Sextans-like galaxy. Indeed Sextans is one the least massive classical dwarfs and it benefits from extensive medium and high resolution spectroscopy, including abundance ratios over a wide metallicity range. This allows for a proper and detailed comparison with the models.


As compared to \citet{Revaz2018}, the supernova efficiency is increased by a factor 2, from $\epsilon=0.1$ to $0.2$. With this
setting, our Sextans-like model \texttt{h177}  has
a total V-band luminosity of about $3\cdot 10^5\,\rm{L_\sun}$, only 25\% lower than the observed value \citep{McConnachie_2012}.  Star formation is fully stopped  after $3 \,\mathrm{Gyr}$, in agreement with the star formation history derived from colour magnitude diagrams \citep{lee2003,lee2009,bettinelli2018}.
The peak metallicity ($-1.97$) and the velocity dispersion  ($\sim 6\,\rm{km/s}$) 
are both in agreement with the values derived from the Calcium triplet \citep{battaglia2011}.


In a $\Lambda$CDM cosmological framework, dwarf galaxies, and particularly UFDs, are dominated by metal-poor stars, which form in clumps in the initial mini-halos. After a few hundred $\mathrm{Myrs}$, these clumps merge and form the final galaxy. This sequence is shown in Fig.~\ref{fig:TimeEvolv}, which displays the build-up history of \texttt{h177} from $190$ to $430\,\rm{Myr}$, a period which corresponds to the first intense phase of the star formation  in this system. For illustration purpose, we show here the model \texttt{PopII+PopIII(CCSNe+FSNe)}, which includes Pop III stars, but the sequence is identical in all models. The surface density of the gas is color-coded in blue. The Pop III and Pop II stars are respectively shown in yellow and red symbols. The hierarchical assembly of the dwarf is clearly seen; UFDs follow the same hierarchical pattern. 

From Fig.~\ref{fig:TimeEvolv}, it is easy to understand that, besides the supernovae feedback energy, two other  parameters are  crucial to the evolution of the model dwarfs: the density threshold, $\rho_{\rm SFR,c}$,  above which stars form and the size of the regions over which the supernovae ejecta are distributed. These two parameters are constrained by considering the trend and dispersion of the stellar abundance ratios, considering the $\alpha$-elements Mg and Ca, as a function of metallicity ([Fe/H]).

\paragraph{The star formation density threshold}

In our previous studies, star formation was enabled when the gas density for particle $i$ was above $\rho_{\rm{SFR},i}$ (defined by Eq.~\ref{eq:pressure}). Indeed, while the probability is low, in principle, it is  possible that stars form from low density gas i.e., lower than $1\,\rm{atom/cm^{3}}$. Together with the gas surface density, Fig.~\ref{fig:TimeEvolv}  traces the  metallicity of the gas ([Fe/H]$_{\rm{gas}}$) in orange. This allows to follow the regions which are enriched by the supernovae ejecta.


In the top panel of Fig.~\ref{fig:TimeEvolv}, $\rho_{\rm SFR,c}$ is set to $1\,\rm{atom/cm^{3}}$ ($1.67\times10^{-24}\,\rm{g/cm^{3}}$). Star formation starts at $\sim100\,\mathrm{Myr}$ in mini halos of low gas density.
Because their reservoir of gas is small, only one generation of stars can occur in each of the mini-halos.  Metals ejected by Pop III stars are dispersed over large regions, of up to 
$\sim 500\, \mathrm{pc}\,h^{-1}$ physical radii. Consequently, only a small fraction of these metals is locked into the second generation of stars. Not only this, but the different sampling of the IMF in each of the mini-halos and the fact that mixing very inefficient between them result in large scatter in abundance ratios ([Mg/Fe], [Ca/Fe]) at fixed metallicity, above the dispersion measured in the observations.

In the bottom panel of Fig.~\ref{fig:TimeEvolv}, $\rho_{\mathrm{SFR},c}$ is set to  $5\,\rm{atom/cm^{3}}$  ($8.36\times10^{-24}\,\rm{gr/cm^{3}}$), i.e., five times the value used in the simulation shown in  the top panel. One immediate impact is to delay the onset of star formation by about $\sim200\,\rm{Myr}$, as particles need time to  merge and sufficiently increase the gas density.

Another consequence of the increase of $\rho_{\mathrm{SFR},c}$ is, that being more massive, the gas reservoirs enable the formation of more than one generation  of stars in each of the mini-halos. The ejecta of the Pop III stars increase the metallicity of the gas to about $\rm{[Fe/H]} = -3$, then multiple Pop II stars form from this enriched gas in the same mini-halo, inheriting the chemical imprints of the Pop III stars. 

Because star formation is more spatially concentrated and the number of gas particles higher, the dilution of the metals  in the ISM is reduced.
Thus, in order to avoid the dispersion of metals in a large unrealistic volume, such as occurring with $\rho_{\mathrm{SFR},c}=1\,\rm{atom/cm^{3}}$, we set the star formation density threshold to $5\,\rm{atom/cm^{3}}$.  This choice results from our investigation of a range of values, from $2$ to $10\,\rm{atom/cm^{3}}$, and best reproduce the observed  scatter in [Mg/Fe] and [Ca/Fe].

\paragraph{Spatial extent of the ejection of metals}



As mentioned above, the extent of the region polluted by the supernova ejecta impacts the chemical abundance pattern of the gas particles, later imprinted in the stars. This includes  metallicity and  abundance ratios.
In the following, we investigate two ejection schemes, with the aim to reproduce the observed galaxy metallicity distributions as well as the trends and scatter of the $\alpha$-element abundance ratios with metallicity.

\begin{figure}[h]
  \centering
  \includegraphics[width=0.49\textwidth]{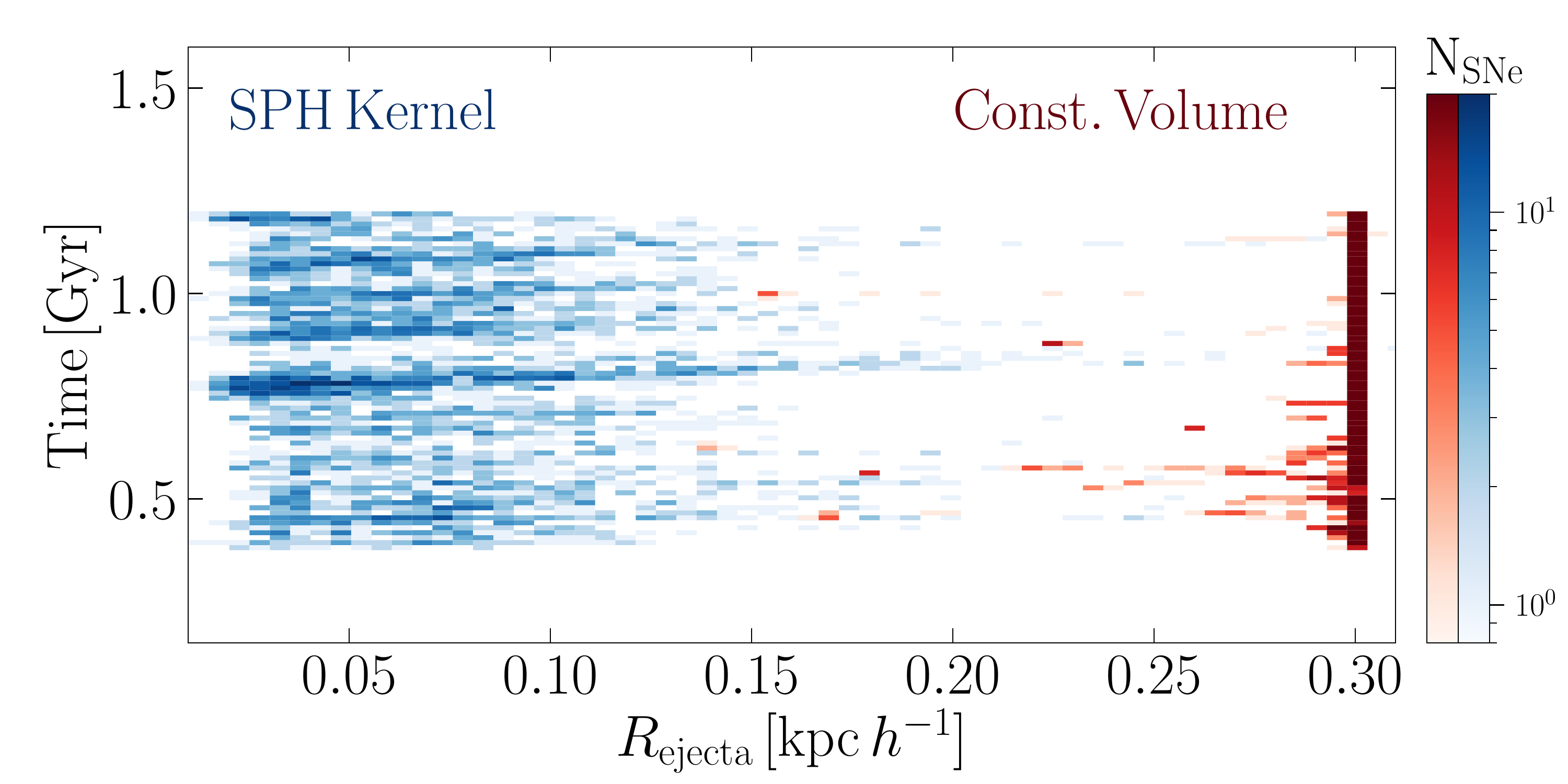}
  \caption{\small Comparison between the two different methods for metal ejection implemented in \textsc{GEAR}. Injection into the \texttt{SPH} kernel and into a constant volume as described in Sec.~\ref{sec:ejection}, are shown in blue and red, respectively, for Sextans-like model \texttt{h177}. Both models are plotted as the 2D histogram of the extent of the polluted region $R_{\mathrm{ejecta}}$, and the time at which the explosion occurs, limited to the first $\sim 1\,\mathrm{Gyr}$ of the simulation. Pixels are color-coded by the number of supernova events. In each event, $R_{\mathrm{ejecta}}$ is defined as the distance to the furthest particle receives the supernova remnants.}
  \label{fig:R}
\end{figure}

We perform a statistical analysis of the properties of the regions reached by the supernovae explosions. The size of the polluted region $R_{\mathrm{ejecta}}$ is defined as the distance to the furthest particle, which receives the supernova ejecta. Fig.~\ref{fig:R} presents the 2D histogram of $R_{\mathrm{ejecta}}$ and time, during the first 1.2 Gyr of evolution of the halo \texttt{h177}. Each point of the diagram corresponds to the number of supernova explosions, which at a given time have enriched the ISM up to $R_{\mathrm{ejecta}}$. Blue  corresponds to the case where the metals are ejected into the \texttt{SPH} kernel, while red stands for the case where the ejecta are distributed in a volume of  constant \textit{maximum} radius of $300\,\rm{pc}$ (see Sec.~\ref{sec:ejection} for the details of both methods). 

In a fully unconstrained way, the \texttt{SPH} radius is generally smaller than $100\,\rm{pc}$, with consequence of leaving a large part of the gas particles in their pristine state, slowing down the  metal-enrichment of the ISM. Therefore, from the point of view of the galaxy mean metallicity, increasing the diffusion radius of the metals is more efficient.

As to the abundance ratios, the general trend of [$\alpha$/Fe] vs [Fe/H] is successfully reproduced by both implementations,  in particular the position of the so-called knee, which indicates the time when the SNeI ejecta significantly contribute to the metal content of the ISM.

%
{}
%


Generally the size of the polluted regions are much smaller in the \texttt{SPH} kernel ejection scheme than in the constant volume method.
In dense gas clouds, where the  $\bar{N}_{\rm{ngb}}$ neighboring particles are found at small  distances, $R_{\mathrm{ejecta}}$ may be as small as $10\,\mathrm{pc}\,h^{-1}$.  
As expected, in the constant volume ejection scheme, the size of the maximum polluted region is  $300\,\rm{pc}$ on average.  This radius can occasionally be smaller in cases of very isolated gas clouds. 

\begin{figure}[h]
    \centering
    \includegraphics[width=0.49\textwidth]{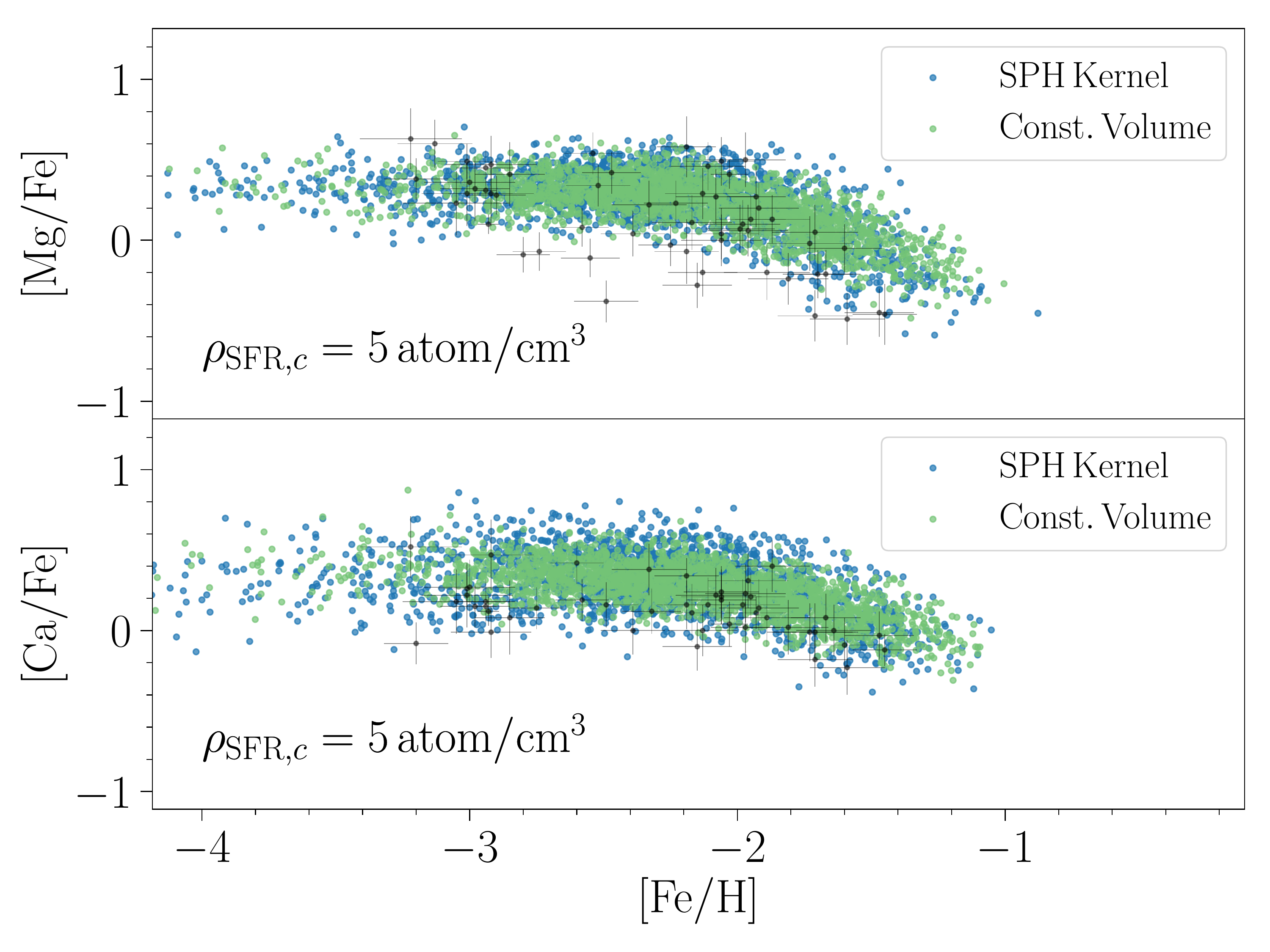}
  \caption{\small [Mg/Fe] and [Ca/Fe] versus $[\rm{Fe/H}]$ for stars formed in the two different methods for metal ejection, compared to the observed stars. 
  Ejection into the \texttt{SPH} kernel and into a constant volume is shown in blue and green, respectively. 
  Observational values for Sextans stars are shown with black dots and error bars. The data are taken from 
\citet{Shetrone_2001}
\citet{Kirby_2010}
\citet{tafelmeyer2010}
\citet{theler2020}
and
\citet{lucchesi2020}. 
  }
  \label{fig:MgCa_sextans_ejecta}
\end{figure}

The small ejecta radii induced by the \texttt{SPH} scheme leads to less efficient metal mixing and consequently larger scatter in the abundance ratios of $\alpha$-elements, as seen  in  Fig.~\ref{fig:MgCa_sextans_ejecta}, which displays the stellar [Mg/Fe] and [Ca/Fe] as a function of $[\rm{Fe/H}]$.  The two settings for metal ejection, i.e., the \texttt{SPH} kernel and the constant volume (see Tab.~\ref{tab:params}), are shown in blue and green, respectively. For the sake of a fair comparison with the observations, we add random uncertainties to the model abundances, following a normal distribution with a standard deviation of $0.1$ dex.  Using the same yields, \citet{kobayashi2020} semi-analytical models produce a [Mg/Fe] plateau $0.2$ dex higher than the observations.  This small difference is interpreted as uncertainties in the stellar yields. We therefore  shift our model [Mg/Fe] by $-0.2$ dex in all subsequent figures involving abundance ratios.
Fig.~\ref{fig:MgCa_sextans_ejecta} shows that the scatter induced by the \texttt{SPH} scheme leads to larger scatter than observed\footnote{Note that the dispersion in the observed [Mg/Fe] is larger than [Ca/Fe] simply because the number of lines used for the derivation of the magnesium is much smaller than for the calcium.} We therefore adopt a constant $R_{\mathrm{ejecta}}$ for our simulations as it leads to an improved  chemical mixing.


%
{}

\subsection{New yields for Pop II stars}

\begin{figure}[h]
  \centering
  \includegraphics[width=0.49\textwidth]{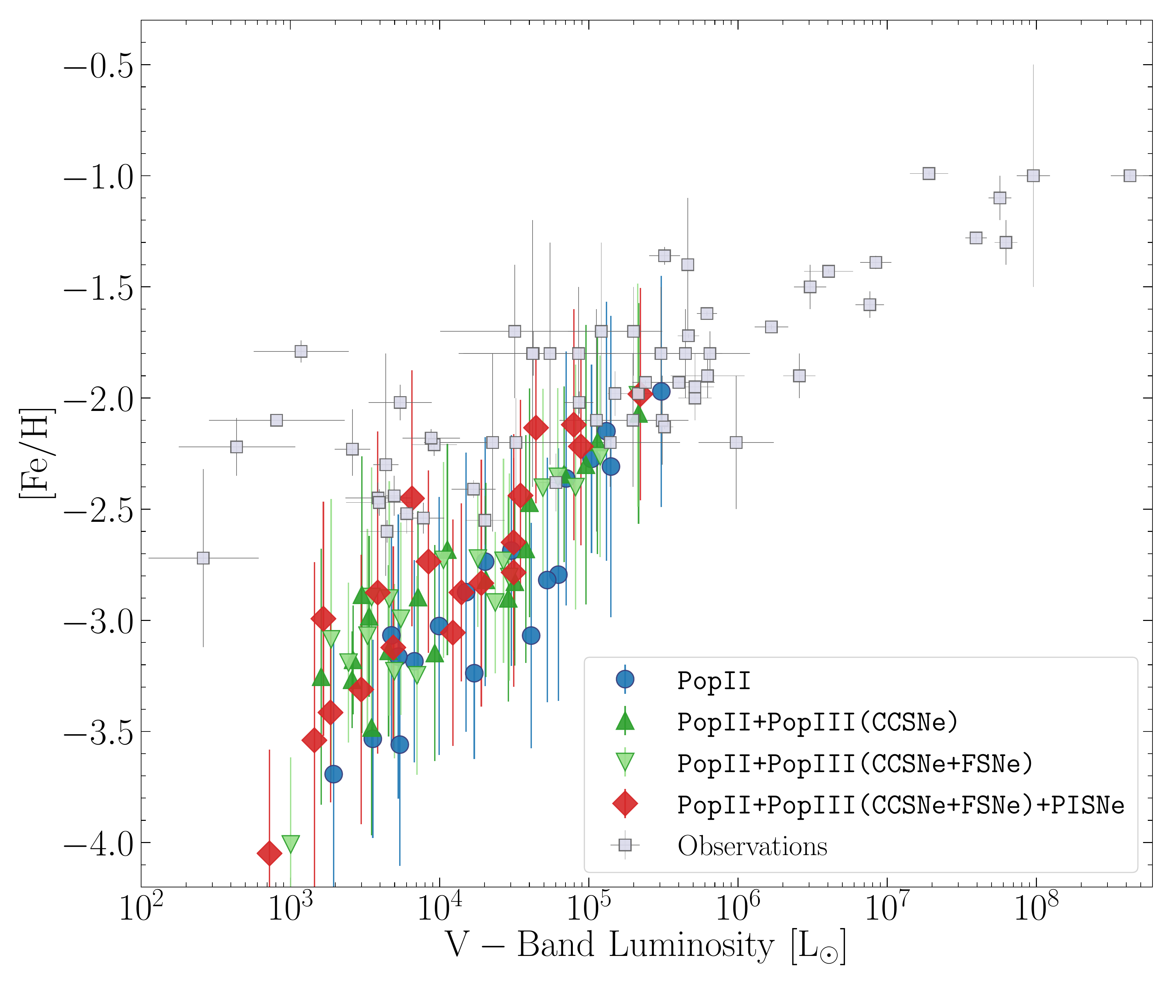}
  \caption{\small Comparison of the luminosity-metallicity relation for dwarfs and UFDs, between Local Group observations and simulations. The grey squares represent our Local Group sample (see text for details). Coloured points stand for different simulations described in Tab.~\ref{tab:params}: The reference \texttt{PopII} model is in blue points, model \texttt{PopIII(CCSNe)} including yields from both massive and low mass Pop III stars is in dark green triangles, model \texttt{PopIII(CCSNe+FSNe)} including yields from only low mass Pop III stars is in light green triangles, and model \texttt{PISNe} including yields  from PISNe is in red diamonds.   
  Error bars are computed following the method discussed in Section~\ref{sec:obs}.
  }
  \label{fig:ZL2}
\end{figure}

Compared to our previous work the resolution, the yields and the injection scheme
have changed. In a first step, we simulate our galaxy models (see Tab.~\ref{tab:catalogue}) 
considering only Pop II stars and check the impact of our updates on the luminosity-metallicity
relation.
Our new galaxy models (\texttt{PopII}) are displayed in blue on Fig.~\ref{fig:ZL2}. 
Compared to the old ones (shown also in blue circles in Fig.~\ref{fig:ZL_base}), the differences are weak.
With the new simulation setting our Sextans model \texttt{h177} ($L_{\mathrm{V}} = 3\cdot10^5\,\mathrm{L}_{\odot}$, $\mathrm{[Fe/H]}=-1.97$), is perfectly superimposed with its counterpart simulated in \cite{Revaz2018}.
At low luminosities, in the UFDs regime, the strong deviation with respect to the observations 
remains. 
Our faintest dwarfs with luminosities of a few $10^3\,\rm{L_{\odot}}$
have a $[\rm{Fe/H}]$ value as low as $-3.5$, about one dex below the observed relation.
It is  worth noting that the new simulation setting slightly reshuffles the luminosity and 
metallicity of our faintest dwarfs when studied individually, on a one to one basis. This underlines the great sensitivity of those objects to any kind of perturbations. Similar changes would have been expected by, for example, a modification to the random seed used in the stochastic star formation recipe \citep[see e.g.][]{Revaz2016,keller2019}. 

From this first experiment, it stands out that using a higher resolution,
but also an updated injection scheme and new yields for Pop II stars do not solve the under-prediction of stellar metallicity at low luminosity. This first set of simulations with only Pop II stars
will be considered as our reference model \texttt{PopII} (see Tab.~\ref{tab:params}). 

\subsection{Pop III stars - without PISNe}\label{sec:popiii}

Looking for  the impact of Pop III stars in the evolution of the UFDs, we first consider them up to $140\,\mathrm{M_\odot}$, i.e., excluding PISNe. Pop III stars with initial masses between $13$ and $30\,\mathrm{M_\odot}$ explode as CCSNe. For stars with masses between $30$ to $140\,\mathrm{M_\odot}$, we consider two  possibilities : 
\begin{itemize}
    \item These stars are FSNe, i.e. not releasing any metal nor energy. This is the \texttt{PopIII(CCSNe+FSNe)} model in Tab.~\ref{tab:params} \citep{kobayashi2020}.
    \item  They explode as normal CCSNe with the  nucleosynthesis yields of \citetalias{NK13}. Even if unlikely, this hypothesis is considered
    to check how much the released mass of metals  can be increased in this stellar mass range.   This is the \texttt{PopIII(CCSNe)} model in Tab.\ref{tab:params}.
   
\end{itemize}

\noindent
The impact of Pop III on the luminosity-metallicity relation  is presented in Fig.~\ref{fig:ZL2}. Light and dark green triangles stand for  the \texttt{PopIII(CCSNe)} and \texttt{PopIII(CCSNe+FSNe)} models, respectively. Two effects explains the shift of the relation: On the one hand, Pop III stars explode in a few $\mathrm{Myr}$ and therefore do not contribute to the final stellar mass budget of their host galaxy. Compared to the \texttt{PopII} only model, the total luminosity is reduced by $\sim$30\%. On the other hand, the IMF of Pop III starts at $13\,\mathrm{M_{\odot}}$, strongly increasing the number of supernovae per unit formed stellar mass. They can all potentially contribute to the metal enrichment. In contrast, in the case of the \texttt{PopII} model, a large fraction is locked in low mass stars ($[0.05-8]\,\mathrm{M_{\odot}}$) which do not participate to the chemical enrichment of their host galaxy.

\begin{table}[h]
\small
  \caption{\small
  Total number of supernova events, number of PISNe, and their corresponding iron ejection mass for the different IMFs used in this work.
  All quantities are computed for $1000\,\rm{M_\odot}$ of stars formed.
  \label{tab:sn}}
  \centering
\begin{threeparttable}
  \begin{tabularx}{0.49\textwidth}{lXXXX}
  \hline
  \hline
  Model ID &  & \#SNe\tnote{1}/ & Fe Total\tnote{3}/     & Fe Boost\tnote{5}\\
     &     &    \#PISNe\tnote{2}    & Fe PISNe\tnote{4}  &   \\
  \hline
  \multicolumn{4}{l}{\texttt{PopII}} \\
   &  & 7.4/0 & 0.29/0  & 1 \\
  \hline
  \multicolumn{4}{l}{\texttt{PopIII(CCSNe)}} \\
   &  & 33.2/0 & 2.6/0 & 9.0\\
  \hline
  \multicolumn{4}{l}{\texttt{PopIII(CCSNe+FSNe)}} \\
   &  & 26.7/0 & 1.9/0 & 6.6\\
  \hline    
  \multicolumn{4}{l}{\texttt{PopIII(CCSNe+FSNe)+PISNe}} \\
    &  & 23.2/0.83 & 7.7/6.2 & 26.6 \\
  \hline
  \hline
  \end{tabularx}
  \begin{tablenotes}
  \item[1]Total number of supernovae (CCSNe and PISNe) for 1000 $\mathrm{M}_{\odot}$
  \item[2]Number of PISNe for 1000 $\mathrm{M}_{\odot}$ of stars
  \item[3]Total iron mass ejected from CCSNe and PISNe per 1000 $\mathrm{M}_{\odot}$ of stars
  \item[4]Iron mass ejected from PISNe per 1000 $\mathrm{M}_{\odot}$ of stars
  \item[5]Ratio of the total mass of ejected iron in each model compared to the \texttt{Pop II}, which contains no Pop III stars 
  \end{tablenotes}
  \end{threeparttable}
  
\end{table}
To illustrate this point, Tab.~\ref{tab:sn} provides the number supernovae, which explode for $1000\,\rm{M_\odot}$ of star formed,  as well as the amount of iron produced by each type of supernovae.
While on average only 7 CCSNe are present in the \texttt{PopII} model, this number rises to 26 for model \texttt{PopIII(CCSNe+FSNe)} and to 33 for \texttt{PopIII(CCSNe)}. This increase in the number of supernovae boosts the released mass of iron. It is multiplied by  $6.6$  in \texttt{PopIII(CCSNe+FSNe)}  and $9$ in \texttt{PopIII(CCSNe)} as compared to \texttt{PopII}.


While the tension with observations is mitigated, thanks to the decrease in galaxy luminosity and increase in metallicity, [Fe/H] still deviates by one dex, for luminosities below $5\cdot10^4\,\mathrm{L_{\odot}}$. 



\subsection{Pop III stars - including PISNe}\label{sec:PISNe}

We now introduce very massive Pop III stars in our simulations which explode as PISNe.  To do so, we extend the maximal mass of the Pop III IMF from $140$ to $300\,\mathrm{M_\odot}$.
The impact of PISNe on the luminosity-metallicity relation 
is shown by the red diamonds in Fig.~\ref{fig:ZL2}. 
While there are still a few number of faint galaxies that lie slightly below the observed relation, 
with some having $[\rm{Fe/H}]$ $\leq-3$, there is a general improvement, arising from (i) an increase in metallicity and (ii) a decrease in luminosity. 

First, as seen in the top panel of Fig.~\ref{fig:fe_sum}, stars with masses above $140\,\mathrm{M_{\odot}}$ provide the largest mass of iron (up to $100$ times higher than the normal CCSNe). Tab.~\ref{tab:sn} shows that an IMF including PISNe
generates 26 times more Fe than in  \texttt{PopII}  and 3 times more than in \texttt{PopIII(CCSNe)}, with PISNe contributing to about $80\%$ to this amount. 
Second, while Pop III stars dominate the UFD population but do not contribute to the total  final mass buget, the luminosity of the \texttt{PISNe} model is again 0.2 dex lower than that of  \texttt{PopII}, similarly to the \texttt{Pop III} models without PISNe.

\begin{figure}[h]
  \centering
  \includegraphics[width=0.49\textwidth]{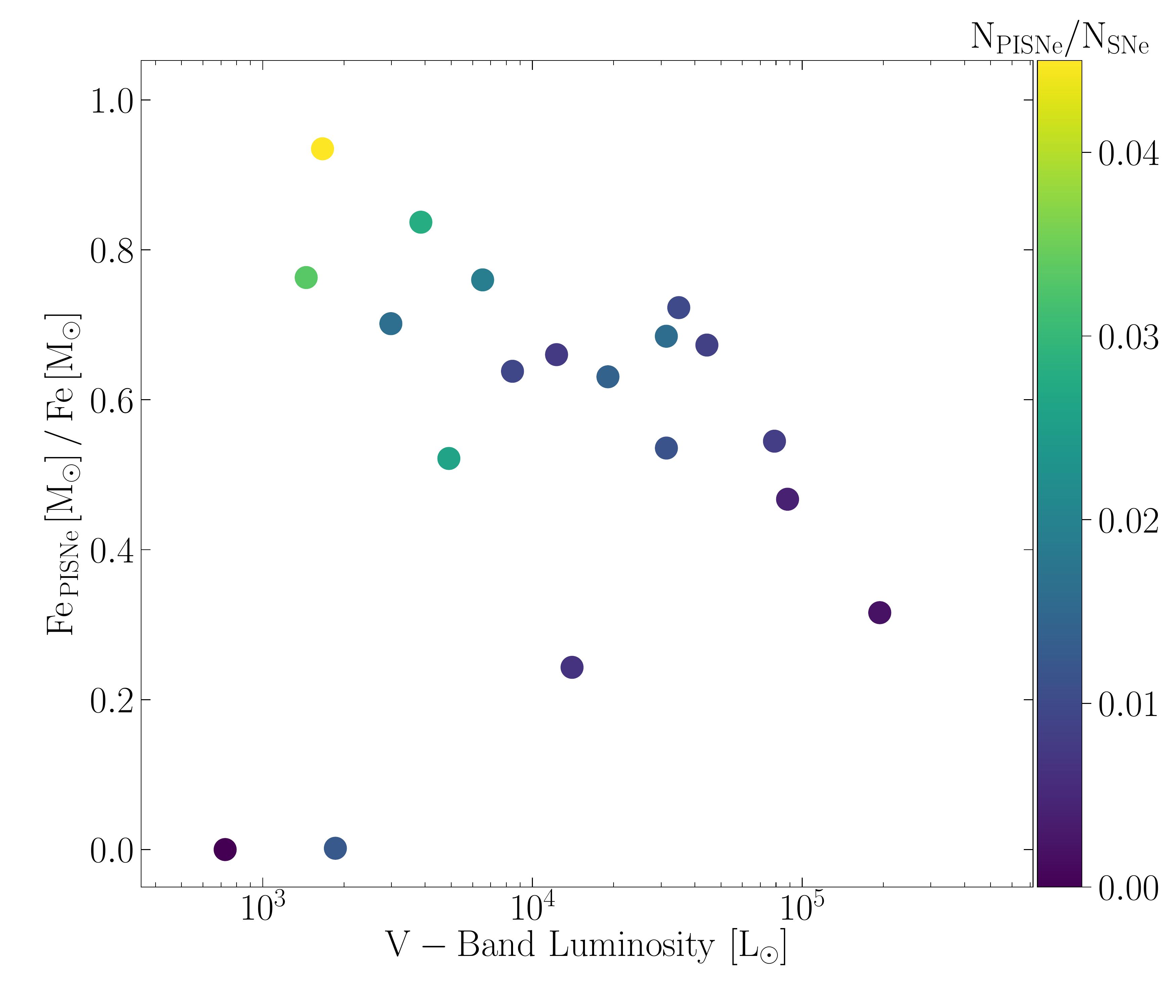}
  \caption{\small The Fe released by PISNe compared to the total Fe produced in all CCSNe and PISNe events. Each point represent one model galaxy color-coded by the ratio between the number of stars forming PISNe and the total number of stars ending up as supernovae. 
  }
  \label{fig:FePISN}
\end{figure}

The higher the fraction of Pop III relative to Pop II stars, the more the above two effects  impact the final luminosity and metallicity of the galaxy. They are particularly significant for the faintest dwarfs, which have shorter star formation histories, as seen in 
Fig.~\ref{fig:ZL2}. However, as PISNe may be extremely rare, even nonexistent in systems forming few stars,
the metallicity is not increased at the faintest end, i.e. at a luminosity of about $10^3\,\rm{L_\odot}$.
 Fig.~\ref{fig:FePISN} quantifies the origin of the iron content of each stellar particle in the \texttt{PISNe} model. In order to do so, the gas particles from which  each of the stellar particle form are identified and their pollution history is analysed. This is done thanks to a logger module which records on-the-fly the time at which a gas particle receives metals from a supernova, allowing 
to distinguish between CCSNe and PISNe sources. 


Fig.~\ref{fig:FePISN} reports on the iron mass fraction originating from PISNe as a function of
galaxy luminosity. Color code the relative number of PISNe. The trend is very clear.
In the faintest systems, up to $4\%$ of supernovae are PISNe, while the same value drops 
below $1\%$ for the brightest. This makes the iron production by PISNe to be dominant (up to 90\%) in the total iron budget of the faint galaxies, while this value drops down to 30\% for a Sextans-like galaxy. Indeed, Sextans and more massive dwarf galaxies are dominated by in-situ star formation.
Thus, after a short period of Pop III star formation, the galaxy is dominated by 
Pop II stars (see Fig.~\ref{fig:Nstars}).  On the contrary, in UFDs, as mentioned in Sec.~\ref{sec:cal}, the first stars
form in isolated small mini-halos. Due to their shallow gas reservoir, only 
few stars can form in the same mini-halo. The gas metallicity stay below the metallicity threshold for the onset of the formation of Pop II stars up to  $1\,\mathrm{Gyr}$. In consequence, Pop III, and in particular PISNe,  have a strong impact on the metallicity and luminosity of the faintest UFDs.

Two exceptions stand out of this general rule: the halo \texttt{h291} with 
$L_{\mathrm{V}} \cong 7\cdot10^{2}\,\mathrm{L}_{\odot}$ does not experience any PISNe 
explosion and \texttt{h273} with $L_{\mathrm{V}} \cong 2\cdot10^{3}\,\mathrm{L}_{\odot}$  
which has only one PISNe exploding in an isolated subhalo.

Indeed, for UFDs  with luminosities below $10^4\,\mathrm{L}_{\odot}$, the expected number of 
PISNe is very low (see Tab.~\ref{tab:sn}). This is confirmed by our models that 
on average form less than five PISNe per UFD.  Thus, due to this stochasticity, PISNe induce a large scatter in the luminosity-metallicity relation rather than a shift towards higher $[\rm{Fe/H}]$. Consequently, PISNe cannot be counted as a reliable source for providing the high metallicity observed in these systems. This also underlines the importance of examining a large sample of UFDs, as done in this study.



%
%

\subsection{Abundance ratios}\label{sec:abundances}

%
\begin{figure*}[h]
\centering
\includegraphics[width=\textwidth]{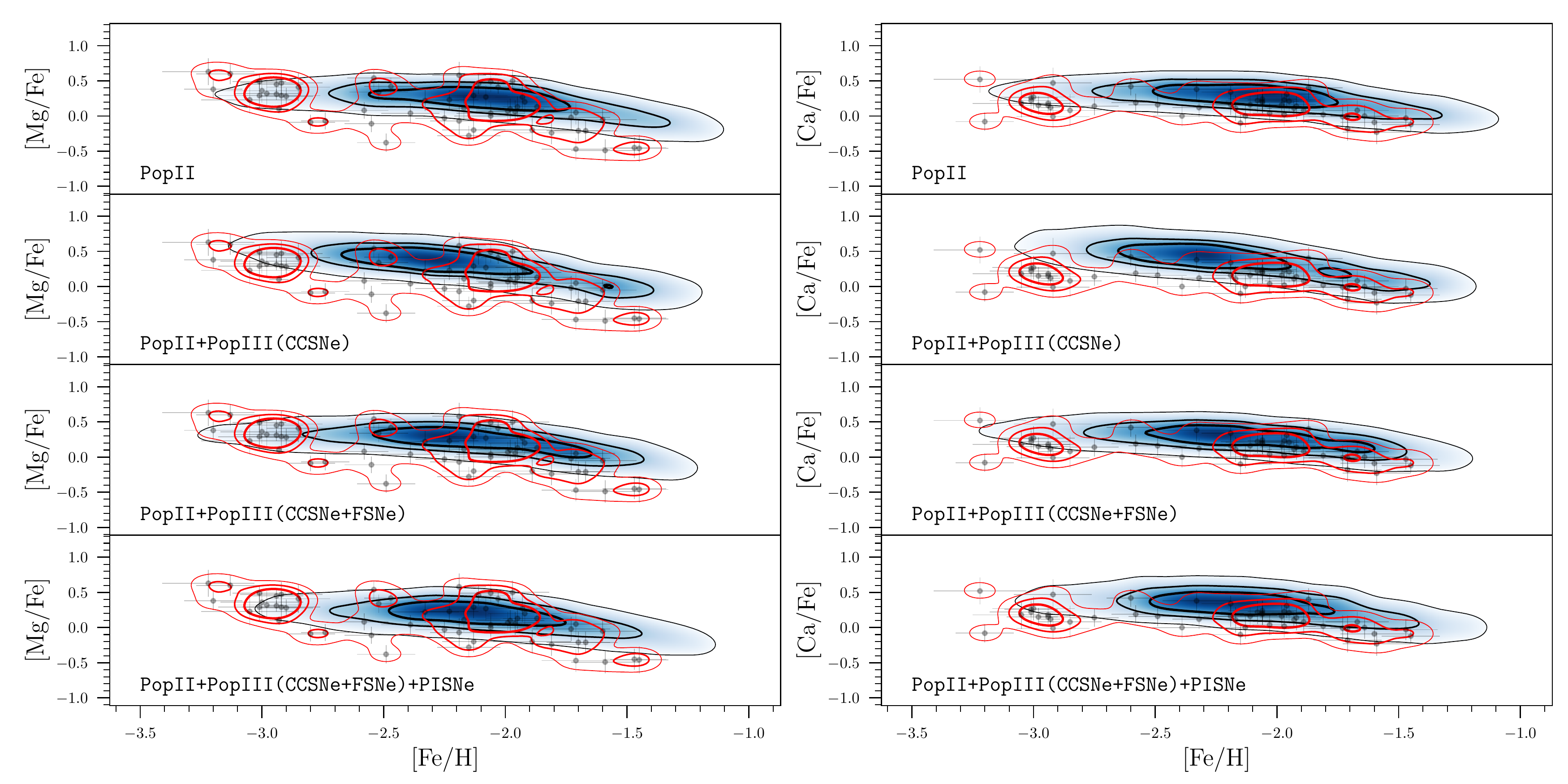}
\caption{\small [Mg/Fe] and [Ca/Fe] abundance ratios as a function of $[\rm{Fe/H}]$ for the dwarf spheroidal model (\texttt{h177}). 
The stellar distribution in blue 
is compared to the Sextans observations, shown in grey points with error bars. 
For both simulations and
observations, we add contours, respectively black and red, that
represent the regions encompassing 30, 60 and 90\% of the stars.
The observational data is the same as in Fig.~\ref{fig:MgCa_sextans_ejecta}.
\label{fig:sextans}}
\end{figure*}
%


Our simulations allow us to assess the impact of the different types of first stars, in more detail than only on the galaxy mean  properties (e.g., [Fe/H], $L_{\mathrm{V}}$), looking at their influence on the chemical abundance ratios of long-lived low mass stars. This is done in comparison with observations, compiling all existing chemical abundance studies with available abundance ratios from high or medium spectroscopic resolution.

\begin{figure}[h]
\centering
\includegraphics[width=0.49\textwidth]{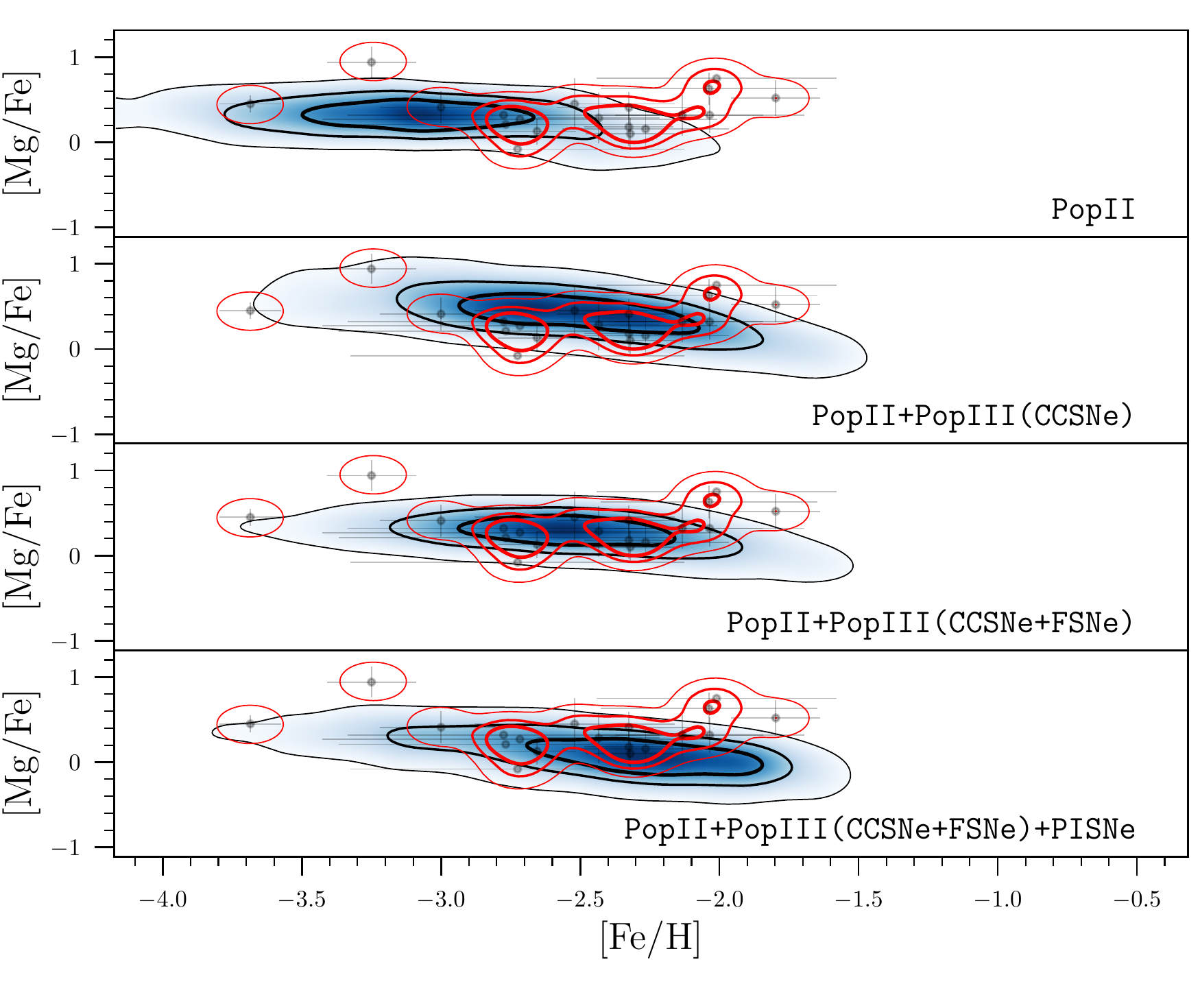}
\caption{\small [Mg/Fe] vs $[\rm{Fe/H}]$ for the UFDs with luminosities larger than 10$^4$ L$_{\odot}$, for the models 
(in blue) and the observations (in gray). 
Contours in black for the models and in red for the observations represent  the regions encompassing 30, 60 and 90\% of the stars.
Data contains a total of 19 stars from
Bootes I \citep{gilmore2013,feltzing2009,norris2010,ishigaki2014}
and
Hercules \citep{koch2008,vargas2013,francois2016}.}
\label{fig:brightUFDs}
\end{figure}
\begin{figure}[h]
\centering
\includegraphics[width=0.49\textwidth]{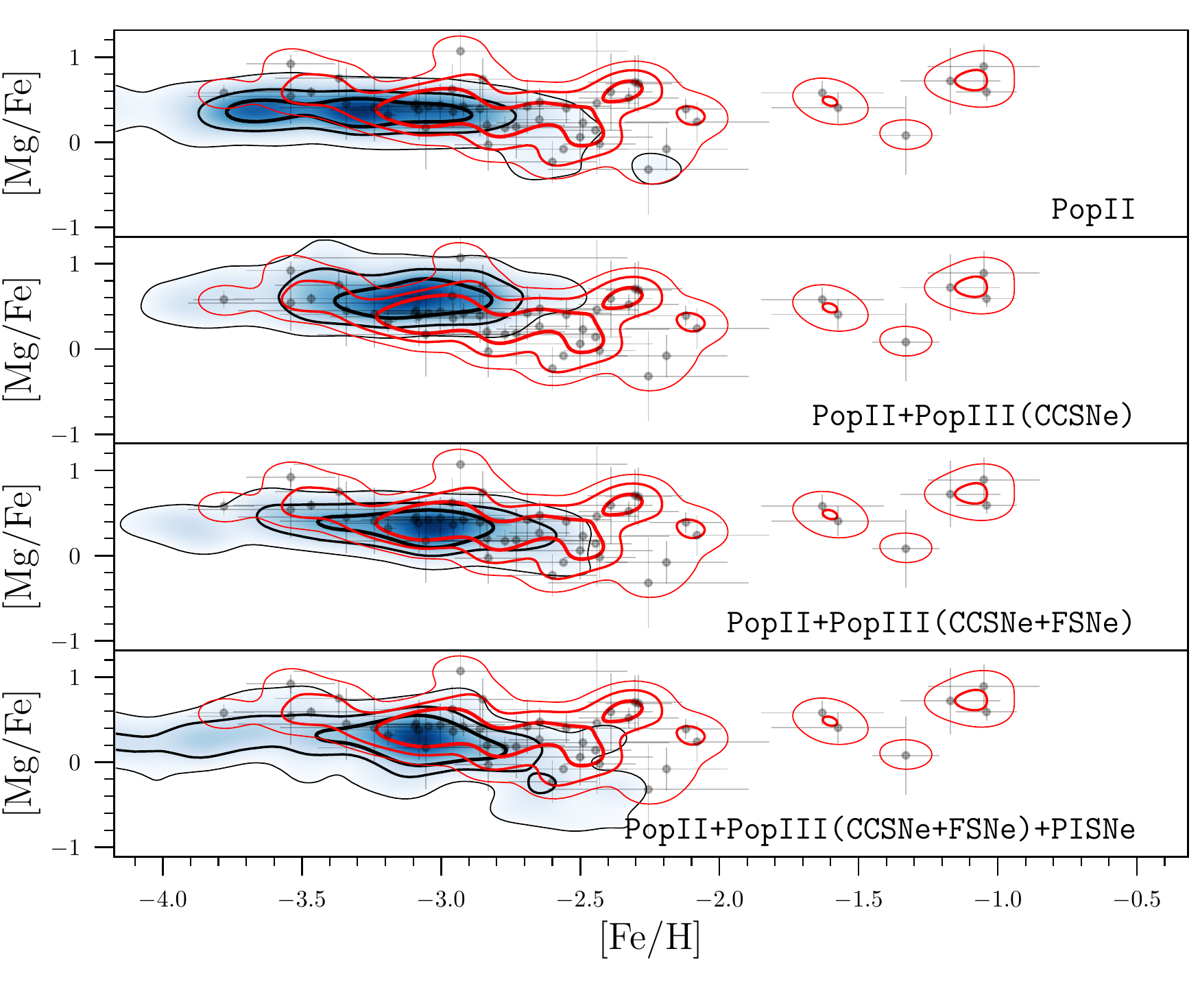}
\caption{\small [Mg/Fe] vs $[\rm{Fe/H}]$ for the UFDs with luminosities lower than 10$^4$ L$_{\odot}$, for the models (in blue) and the observations (in gray). Contours in black for the models and in red for the observations represent the regions encompassing 30, 60 and 90\% of the stars.
Data contains a total of 49 stars from
Bootes II \citep{ji2016c,koch2009,francois2016},
Canes Venatici II \citep{francois2016,vargas2013},
Grus I \citep{ji2019},
Horologium \citep{nagasawa2018},
Leo IV \citep{simon2010,francois2016},
Reticulum II \citep{ji2016,ji2016b,roederer2016,ji2018},
Segue I \citep{norris2010b,vargas2013,frebel2014},
Segue II \citep{kirby2013b},
Triangulum II \citep{venn2017,kirby2017,ji2019},
Tucana II \citep{ji2016d,chiti2018},
Tucana III \citep{hansen2017}
and
Ursa Major II \citep{frebel2010b,vargas2013}. }         
\label{fig:faintUFDs}
\end{figure}

Fig.~\ref{fig:sextans} presents the [Mg/Fe] and [Ca/Fe]
stellar abundance ratios as a function of [Fe/H] for our Sextans model (\texttt{h177})
and different prescriptions listed in  Tab.\ref{tab:params}. 
These two $\alpha$-elements are particularly well documented for the Sextans dSph, handling  both  nucleosynthesis and star formation timescales. 
In Figs. \ref{fig:brightUFDs} and \ref{fig:faintUFDs}, for our UFD models, we restrict the comparison with observations to Mg.
It is indeed the  element which has been most often  determined  in these faint systems. However only a handful of stars per each individual UFD benefit from detailed chemical analysis (the maximum being Bootes with 13 stars). Therefore, we draw the comparison of our 18 UFD models with the combination of the existing observations of 68 stars in 14 UFDs. Both simulations and observations are however split
in two luminosity intervals, with $L_{\rm V}>10^4\,\rm{L_{\odot}}$ and 
$L_{\rm V}<10^4\,\rm{L_{\odot}}$.
In Fig.~\ref{fig:sextans}, \ref{fig:brightUFDs}, and \ref{fig:faintUFDs}, the model probability distributions are shown in blue, while the observations are displayed with gray points. The black and red contours delineate the regions encompassing 30, 60 and 90\% of the stars, in the models and observations, respectively.



\begin{figure}[h]
    \centering
    \includegraphics[width=0.49\textwidth]{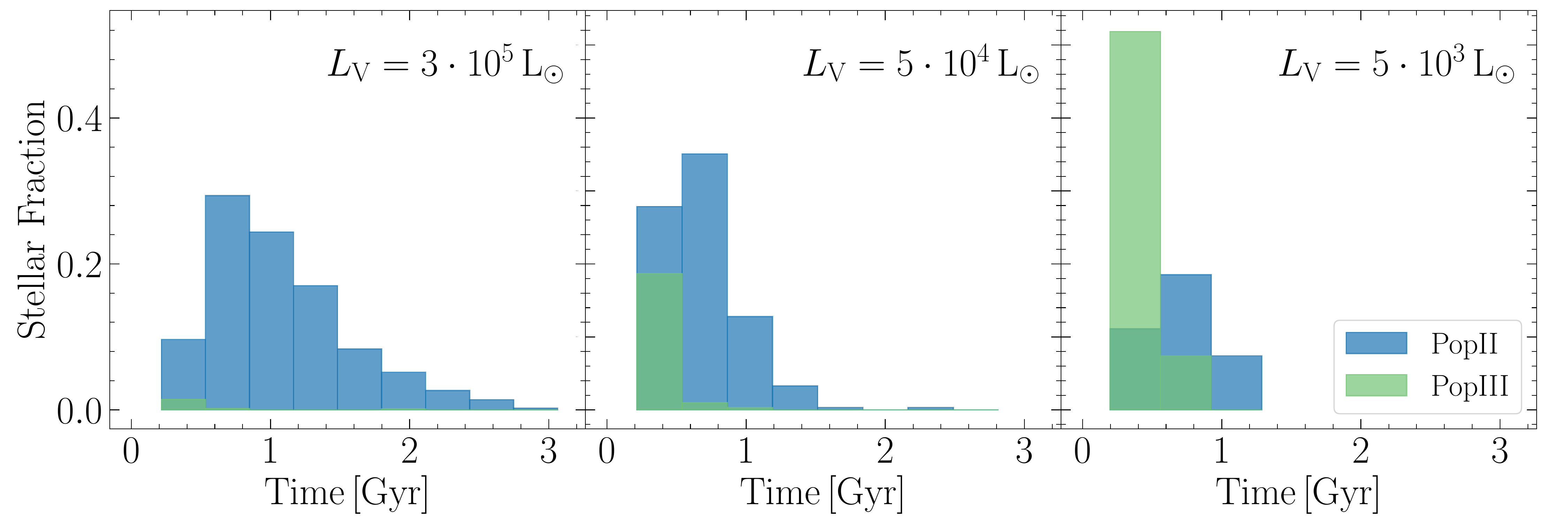}
  \caption{\small Fraction of formed Pop II (in blue) and Pop III (in green) stars. Each panel corresponds to one halo with its total luminosity indicated on the upper right side of the panel. The  left panel shows our Sextans-like model, \texttt{h177}, with $L_{\mathrm{V}} \sim 3\cdot10^5\,\mathrm{L}_{\odot}$. The middle panel shows \texttt{h152} ($L_{\mathrm{V}} \sim 0.5 \cdot10^5\,\mathrm{L}_{\odot}$) representative of medium mass UFDs. The  right panel displays \texttt{h323}, representative of small mass UFDs ($L_{\mathrm{V}} \sim 0.05 \cdot10^5\,\mathrm{L}_{\odot}$ ), with quenched star formation after $1\,\mathrm{Gyr}$.}
  \label{fig:Nstars}
\end{figure}

All models of the \texttt{h177} halo feature the same general trend with increasing metallicity, starting with a super solar [$\alpha$/Fe] plateau below [Fe/H]$\sim -2$, followed by a knee and a descending branch down to sub-solar [$\alpha$/Fe] values, under the influence of the iron-rich Type Ia supernova ejecta. This is the consequence of a relatively extended star formation history, over $3\,\rm{Gyr}$, imprinted in the Pop II stars, which constitute up to $98\%$ of the total stellar populations in these massive systems (as shown in Fig.~\ref{fig:Nstars}). 
However, some differences between the models are noticeable, which stem from the specific nucleosynthesis of each type of Pop III stars, but also the total number of supernova explosions and feedback energy.  In particular, keeping as reference the \texttt{PopII} model, the highest concentration of stellar particles is located at lower [Fe/H] values, by about 0.2 dex for the \texttt{PopIII(CCSNe)} model. Moreover,  a small but still significant fraction of the population has higher [Ca/Fe] than  in the observations at [Fe/H] $<-2$, suggesting that it is indeed unlikely that stars in the mass range  $30$ to $140\, \mathrm{M_{\odot}}$ contribute to the galaxy chemical enrichment, in agreement with the FSNe mass range of \citet{kobayashi2020}.

Figures \ref{fig:brightUFDs} and \ref{fig:faintUFDs} allow a deeper assessment of the impact of the  Pop III in the evolution of UFDs above and below $10^4\,\mathrm{L}_{\odot}$, respectively. 
In both luminosity ranges, the main change in the chemical evolution of UFDs is the shift of the 
metallicity peak towards higher values, as seen in Fig. \ref{fig:ZL2}. 
For the brightest UFDs, the tail of the distribution of stars at ${\rm[Fe/H]} \le - 3.5$ also disappears.  
Pop III stars remain  influential in these bright UFDs, still accounting for 38\% of all stars formed 
(see Fig.~\ref{fig:Nstars}), but the choice of the type of Pop III  has little influence on the position of 
the [Fe/H] peak and the distribution in both [Mg/Fe] and [Fe/H] around it.   
Indeed, in these systems, the global galaxy evolution is dominated by the Pop II stars. This explains the 
limited sensitivity to the specific nature (CCSNe, FSNe or PISNe) and mass range
of the first stars. 
Though PISNe, which in average produce less magnesium than iron, induce a  
decrease in [Mg/Fe] and the formation of metal-rich $\alpha$-poor stars, accentuating the subsolar [Mg/Fe] 
branch to a level that does not appear to be supported by the observations. 


The faintest UFDs, for which Pop III stars contribute to 86\% of the full population (see Fig.~\ref{fig:Nstars}), are more sensitive to the nature of the first stars, with the [Fe/H] distribution becoming slightly more peaked.
The  tail of the [Fe/H] distribution below $-3.5$ remains, irrespective of the Pop III nucleosynthesis prescription.
More importantly, the intrinsic chemical evolution of these systems hardly exceeds ${\rm[Fe/H]} \sim -2.5$, contrary to the observations, which seem to extend much beyond.
Given the extremely short star formation history of these systems, it is very difficult to see how their intrinsic evolution alone could ever reach metallicities as high as $-1$. Further comparison with the observations, based on the metallicity distribution is conducted in Sec. \ref{CaT}.


\subsection{Robustness of the results}



We check hereafter the robustness of our results with respect to different model parameters, which could potentially affect our conclusions,  specifically 
the metallicity threshold $\rm{[Fe/H]_c}$, the feedback energy of PISNe, which are uncertain. Finally, we look into the comparison of the observed and model metallicity distributions, in order to help provide deeper insight on the resisting shift between their corresponding $[\rm{Fe/H}]$.



%
%

\subsubsection{Metallicity threshold $\rm{[Fe/H]_{c}}$}\label{sec:MetallicityThreshold}

%
\begin{figure}[hb]
  \centering
  \includegraphics[width=0.49\textwidth]{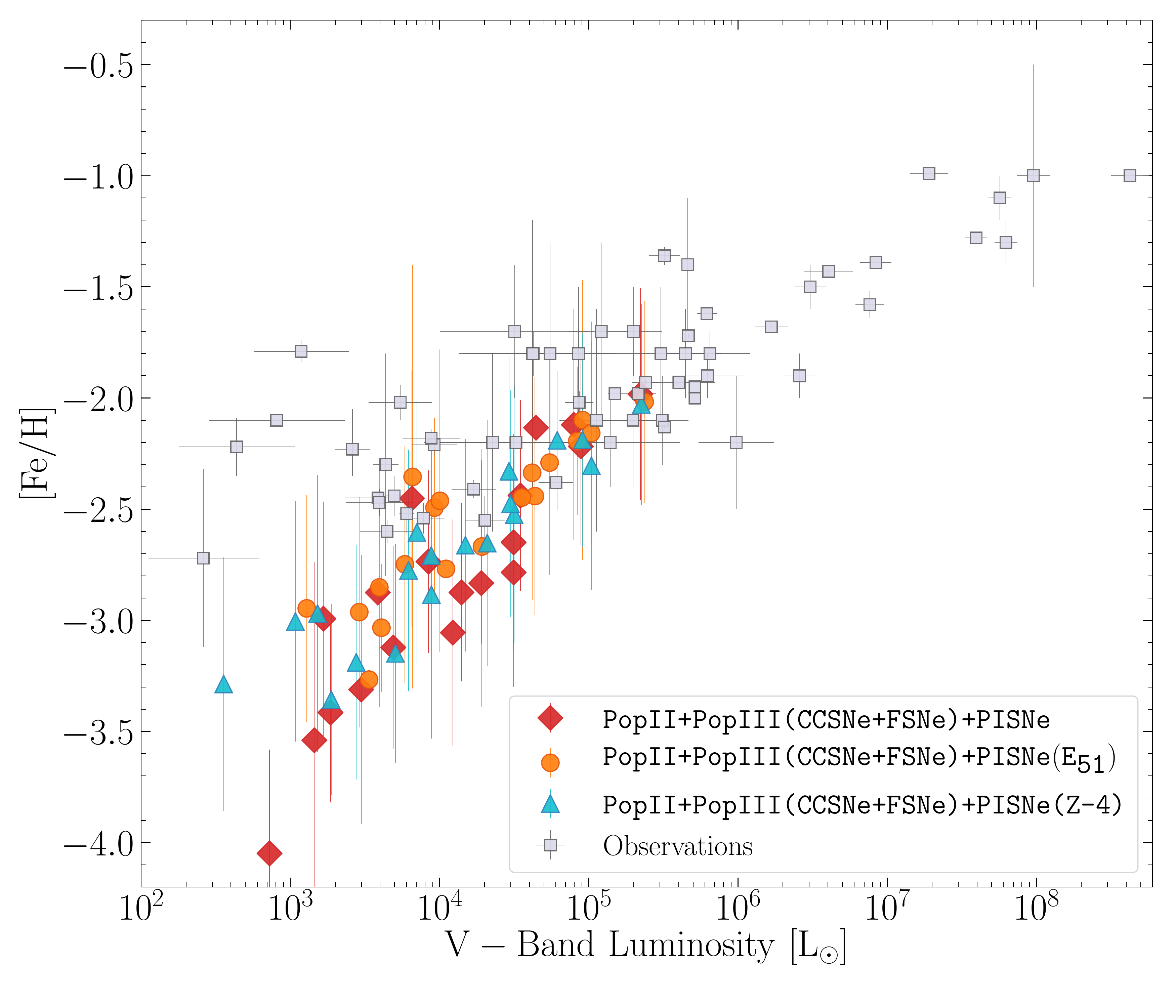}
  \caption{\small Tests of different prescriptions. The \texttt{PISNe} model is shown in the same red diamonds as in Fig.~\ref{fig:ZL2}. Models with a fixed value of $1\cdot10^{51}\,\mathrm{erg}$ are shown in orange circles. Models with a critital metallicity of $-4$ are displayed with blue triangles.}
  \label{fig:LZ-EM}
\end{figure}

We re-simulate the \texttt{PISNe} model, increasing  $\rm{[Fe/H]_c}$ from the 
fiducial value of $-5$ to $-4$.
Increasing $\rm{[Fe/H]_c}$  leads to an increased number of Pop III stars
and consequently a potential increase of the final metallicity of UFDs, given that Pop III stars can provide
up to three times more iron than Pop II stars (Sec.~\ref{sec:popiii}).
Fig.~\ref{fig:LZ-EM} compares the final metallicity of the $\rm{[Fe/H]_c} =-5$ (red diamonds) and $\rm{[Fe/H]_c} =-4$ (blue triangles) models as a function of metallicity. For the vast majority of the models, the impact is very limited.
%
%
It is the strongest for the faintest systems such as \texttt{h291} and \texttt{h273}, because the  probability of PISNe explosions is very low in these low mass systems (Sec.~\ref{sec:PISNe}), the addition of even one single event is very significant. Nevertheless,  increasing $\rm{[Fe/H]_c}$ does not lead to the monotonic upward shift of the global $L_{\mathrm{V}}$-$\rm{[Fe/H]_c}$ relation, which would make them overlap.

\begin{figure*}[htb]
  \centering
  \includegraphics[width=\textwidth]{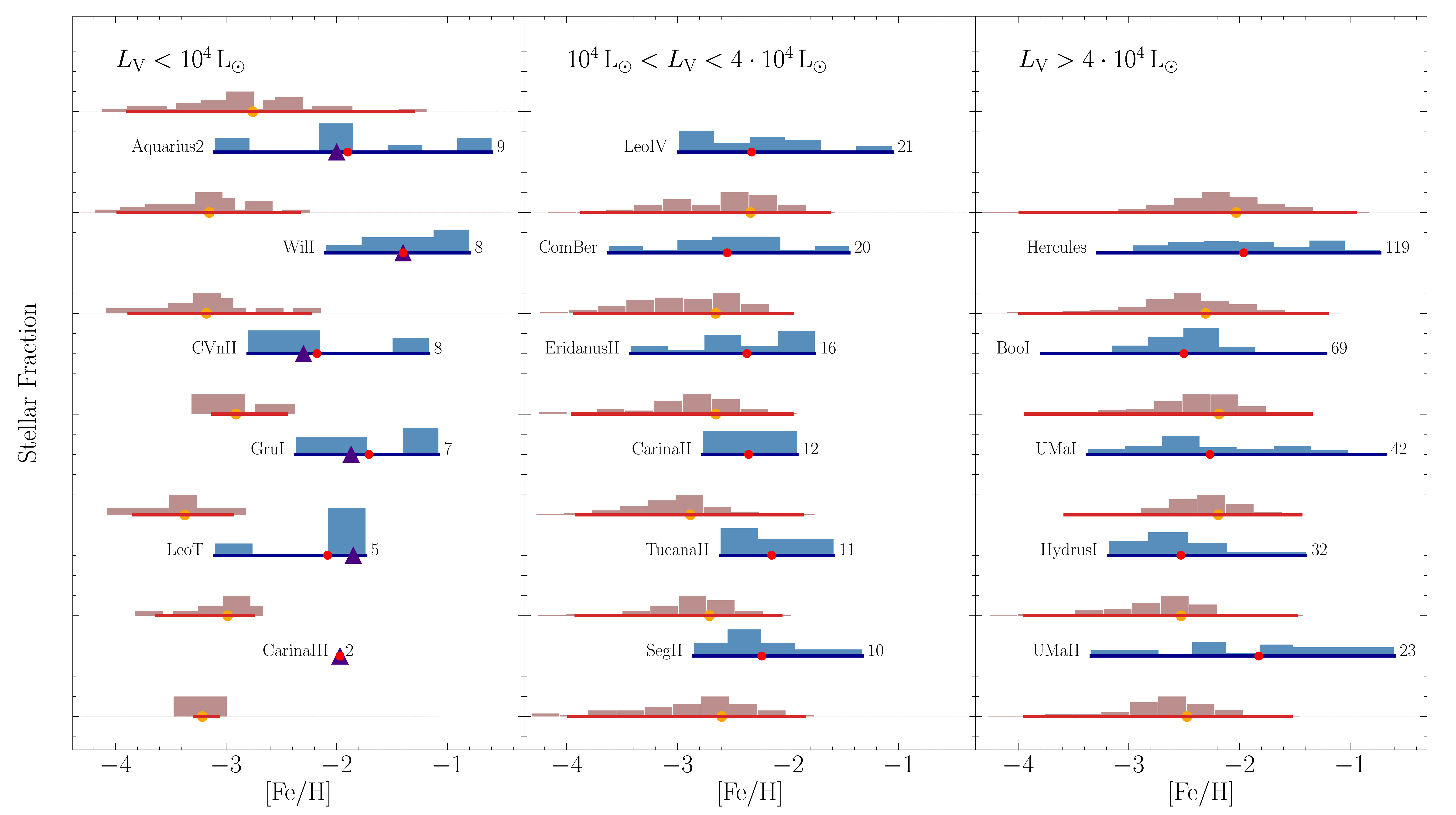}
  \caption{\small Comparison of the metallicity distribution functions (MDF) of the \texttt{PISNe(Z-4)} model (in red) and a set of observed galaxies (in blue) in three  luminosity ranges. 
  The solid red and blue horizontal lines indicate the full extent of the stellar metallicity distributions in the simulations and observations, respectively. The orange points show the mode of the $[\rm{Fe/H}]$ model distribution (see Appendix~\ref{sec:appendix1}). The purple triangles and red points represent the median and mean of metallicities for each observed galaxy. The number of the spectroscopically confirmed member stars with metallicity estimates is indicated.            
  }
  \label{fig:obs}
\end{figure*}

\subsubsection{PISNe energy}

As introduced in Sec.~\ref{sec:PISNe},  the energy of PISNe 
strongly increases with the progenitor mass, from 9 to 50$\cdot10^{51}\,\rm{erg}$  (see Fig.~\ref{fig:fe_sum}).
Between 0 and 5 PISNe can explode in faint UFDs ($L_{\mathrm{V}}<10^4\,\mathrm{L_{\odot}}$), and up to 15 in brighter ones. Thus, in extreme cases, more than $10^{53}\,\rm{erg}$ can  be released in the ISM, limiting the star formation activity in the early stages of the galaxy evolution. 
%
We re-simulate the \texttt{PISNe} model, decreasing the energy of PISNe explosions  to a constant value of $10^{51}\,\rm{erg}$, corresponding to the standard CCSNe feedback energy.  
In Fig.~\ref{fig:LZ-EM} the orange circles represent the \texttt{PISNe($\texttt{E}_{\texttt{51}}$)} 
model with $10^{51}\,\rm{erg}$ feedback, while the red diamonds show our 
fiducial \texttt{PISNe} model. The net effect of the reduction of the feedback energy is that there is no system below $10^{3} \mathrm{L_{\odot}}$ formed anymore, most likely because star formation was less hampered by the PISNe explosions.  Similarly to increasing  $\rm{[Fe/H]_{c}}$, while the metallicities slightly increased at fixed luminosity, this is not in a sufficient large amount to make the observed and modeled relations match.

\subsubsection{UFD metallicity distribution functions}
\label{CaT}


The challenge of reproducing the UFDs calls for an even closer look at the structure of the models but also at the status of the observations and in particular at how well they sample the properties of these galaxies. To this end, Fig.~\ref{fig:obs} compares the metallicity distribution function (MDF) of the \texttt{PISNe(Z-4)} model, chosen because it is the closest to the observed metallicity-luminosity relation in Fig.\ref{fig:LZ-EM},  with the observed distributions derived from the calcium triplet, in three  luminosity intervals.  The stellar samples are taken from the references in \citet{Battaglia2022} and \cite{McConnachie_2012}.



The observed MDF of each  galaxy is displayed in blue. The number of spectroscopically confirmed  galaxy members used in these MDFs is provided. The blue horizontal lines indicate the full metallicity ranges.  The red points locate the mean $[\rm{Fe/H}]$ values. For the faintest systems, with less than 10 member stars, we indicate the median $[\rm{Fe/H}]$ value with purple triangles. Although the number of member stars with available spectroscopic metallicities is less than 20 in most of UFDs, it is already clear that they show a large variety of properties and a wide spread in metallicities up to more than 2 dex.  The MDFs  of the simulated galaxies are shown in red and their full metallicity ranges are indicated  with red solid lines. The MDF peaks are seen with orange points. They are calculated in Appendix~\ref{sec:appendix1}.

As expected, the number of spectroscopically confirmed members decreases with the luminosity of the galaxies, with the consequence that the metallicity distributions may not be fully sampled. They are in any case  very fragmented, making the position of their mean values rather uncertain.  Less expected is that the low metallicity tail, at ${\rm[Fe/H]}\le -3$, seems to gradually disappear towards  less massive systems. Conversely, the high metallicity tail of the distributions remains.  The question arises as to whether this apparent bias is simply a reflection of incomplete sampling or of different intrinsic evolution. In contrast, by their very nature, the models include all stellar populations. They keep the majority of the metal-poor population regardless of the mass/luminosity of the galaxy, whereas the full range of metallicity covered depends on the extent of the star formation history. As commented in the previous section, there are few stellar particles above ${\rm[Fe/H]}= -2$, and only in one case do we reach  ${\rm[Fe/H]}= -1$. 
In summary, it appears from this analysis that we are facing two limits, one observational, with the sampling of UFD properties still partial, but also a numerical limit with the challenge of reproducing systems that, despite very short star formation histories, yet contain stars as chemically enriched as in more massive systems.

\section{Conclusions and Discussion}\label{sec:conclusions}

%


Despite their many successes of recent years, the cosmological simulations of dwarf galaxies still fail at the smallest scales.  Specifically, in the UFD regime,  hydrodynamical models predict a steeper metallicity-luminosity relation than the one observed. Below $10^5\,\rm{L_{\odot}}$, the galaxy mean metallicities can be as much as two dex too low at given luminosity. In order to tackle this issue and understand how very faint galaxies acquire their metals,
we studied the impact of Pop III stars on the evolution of UFDs. To this end, we performed cosmological zoom-in chemo-dynamical simulations from redshift 70 to zero of nineteen halos hosting  dwarfs with V-band luminosities between $2\cdot 10^3$ to $3\cdot 10^5\,\rm{L_\sun}$. The most massive of these halos is a Sextans-like spheroidal dwarf galaxy, while  the eighteen others are fully in the UFD regime,  with luminosities equal or below $10^5\,\rm{L_\odot}$. 
We validated our models by reproducing both the global properties and the $\alpha$-elements trend of the Sextans dwarf galaxy.
We have checked the robustness of our results with respect to model-dependent parameters or poorly constrained physics. 
In particular, we have looked into the effect of the critical metallicity $\rm{[Fe/H]}_c$, that defines the transition 
from Pop III to Pop II stars. We also looked into the energy release from PISNe to see how it could affect the UFD build-up 
and physical properties.

Our main conclusions can be summarized as follows:

\begin{itemize}

\item  In a $\Lambda$CDM cosmological framework, dwarf galaxies, and particularly UFDs, are dominated by metal-poor stars, which form in clumps hosted by the initial mini-halos. After a few hundred $\mathrm{Myrs}$, these clumps merge and form the final galaxy.


\item  The evolution of UFDs is strongly influenced by Pop III stars.  On average, 38\% of all stars have been formed from gas at $\rm{[Fe/H]}<-5$. In small mass UFDs, for which star formation is completely quenched after $1\,\mathrm{Gyr}$, this fraction rises to 86\%.  

\item 
Assuming that the Pop III stars IMF is a Kroupa one with a  slope of $-1.3$, as usually considered for the massive stars,
the main change in the chemical evolution of the dwarf galaxies due to the contribution of the first stars nucleosynthesis 
is the upward shift of the peak of their final metallicity distributions. Considering such IMF,
the nature of these first stars, CCSNe, FSNe or PSINe, is of secondary importance.

\item Pop III stars with masses lower than $140\,\rm{M_\odot}$  can increase the production of iron by up to a factor of  
6 per $1000\,\rm{M_{\odot}}$ of stars formed. This increase is however still  insufficient to match the observed metallicity-luminosity relation. 



\item 
PISNe, with progenitor masses between $140\,\rm{M_\odot}$ and $300\,\rm{M_\odot}$, increase the production of iron by a factor of 26 per $1000\,\rm{M_{\odot}}$ of stars formed.
However, PISNe are rare events and occasionally absent in the faintest UFDs. Therefore they have a limited impact on the global $L_{\mathrm{V}}$ - $[\rm{Fe/H}]$ relation.   

\end{itemize}

 The analysis of the galaxy metallicity distributions and trends of the stellar particle abundance ratios  as a function of [Fe/H] provides further insight into the impact of Pop III stars. 

\begin{itemize}

\item  For the brightest UFDs, above $L_{\rm V}=10^4\,\rm{L_{\odot}}$, the tail of the distribution of stars at [Fe/H] $\le -3.5$ disappears, while it remains for the fainter ones. None of the  $L_{\rm V} \ge 10^4\,\rm{L_{\odot}}$ model UFDs contain stellar particles above [Fe/H] $= -2.5$, contrary to the observations. It remains a clear challenge to produce very small mass systems whose intrinsic evolution can create stars up to [Fe/H] $= -1$ on  time scales smaller than $\sim 1$ Gyr.

\item 
Because PISNe release less magnesium than iron, low-mass long-lived stars formed from gas enriched by their ejecta can have subsolar $[\rm{Mg/Fe}]$. These seems to be outnumbered at [Fe/H]$\ge -2.5$ compared to the observations of UFDs. 

\item 
The analysis of the model abundance ratios rules out any significant contribution from Pop III stars  in the $30\,\rm{M_\odot}$ and $140\,\rm{M_\odot}$ mass range. Indeed, they predict, for all dwarf galaxies, supersolar [Ca/Fe] and [Mg/Fe] ratios, in contradiction with the observations.

\item  From an observational point of view, the number of known spectroscopically confirmed members, with metallicity estimates and/or detailled chemical abundances, is still limited for all UFDs. It decreases even more with galaxy luminosity. As a  consequence,  the observed metallicity distributions can be still very fragmented, making the position of their mean values uncertain.  Moreover the low metallicity tail, at ${\rm[Fe/H]}\le -3$, seems to gradually disappear towards  less massive systems.

\end{itemize}

\noindent
Challenges to be taken up are thus both observational and numerical, with the need to reproduce systems that, despite very short star
formation histories,  contain stars as chemically enriched as in more massive ones.

\begin{acknowledgements}

%

We are grateful to Nicolas Longeard for kindly sharing with us some of the spectroscopic members used in the metallicity histograms presented in this paper. We would  like to thank
Georges Meynet and Chiaki Kobayashi for very useful discussions.    
We acknowledges the support by the International Space Science Institute (ISSI), 
Bern, Switzerland, for supporting and funding the international team “First stars in dwarf galaxies”. 
This work was supported by the Swiss
Federal Institute of Technology in Lausanne (EPFL) through the use of the
facilities of its Scientific IT and Application Support Center (SCITAS). 
The data
reduction and galaxy maps have been performed using the parallelized Python
\texttt{pNbody} package (\url{http://lastro.epfl.ch/projects/pNbody/}).
\end{acknowledgements}

%
\bibliographystyle{aa}
\bibliography{UFDs}
%
\clearpage
\onecolumn

%

\begin{appendix}

%

\section{Determining $[\rm{Fe/H}]$ for simulated galaxies}\label{sec:appendix1}

Properly attributing a metallicity to each simulated dwarf galaxy is essential in this work.
Knowing the metallicity of each stellar particle, the dwarf metallicity is usually computed
by taking the mean, the median or the mode of the metallicity distribution function (MDF).
By taking the mean or the median, one would however strongly underestimate the system 
metallicity, owing to the presence of stars with very low metallicity in simulations that contain only Pop II.
Taking the mode is more appropriate but the latter may be hardly determined in noisy MDFs. 
Increasing the number of metallicity bins will help however, at the expense of the precision.

\begin{figure}[h]
\centering
\includegraphics[width=0.3\textwidth]{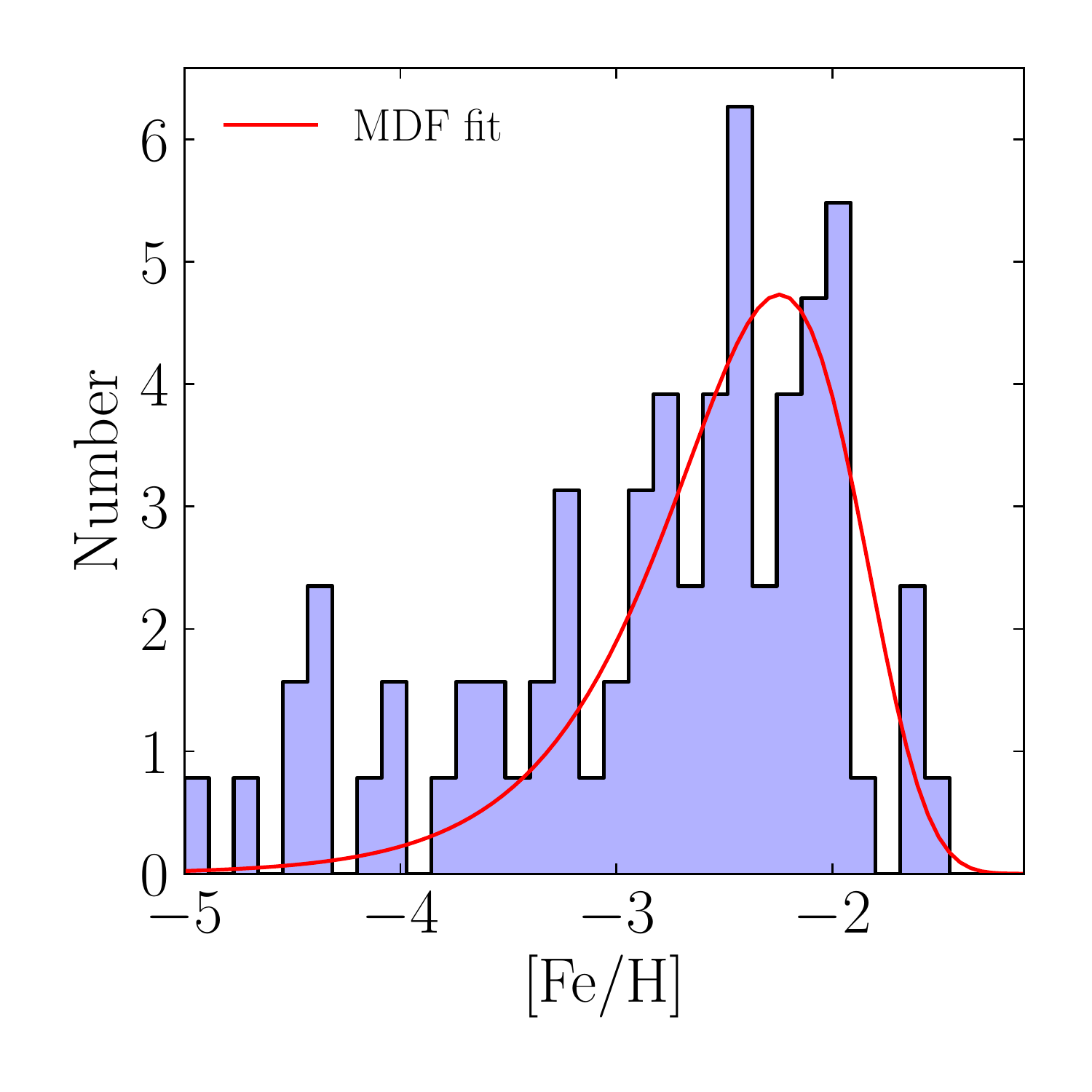} \includegraphics[width=0.3\textwidth]{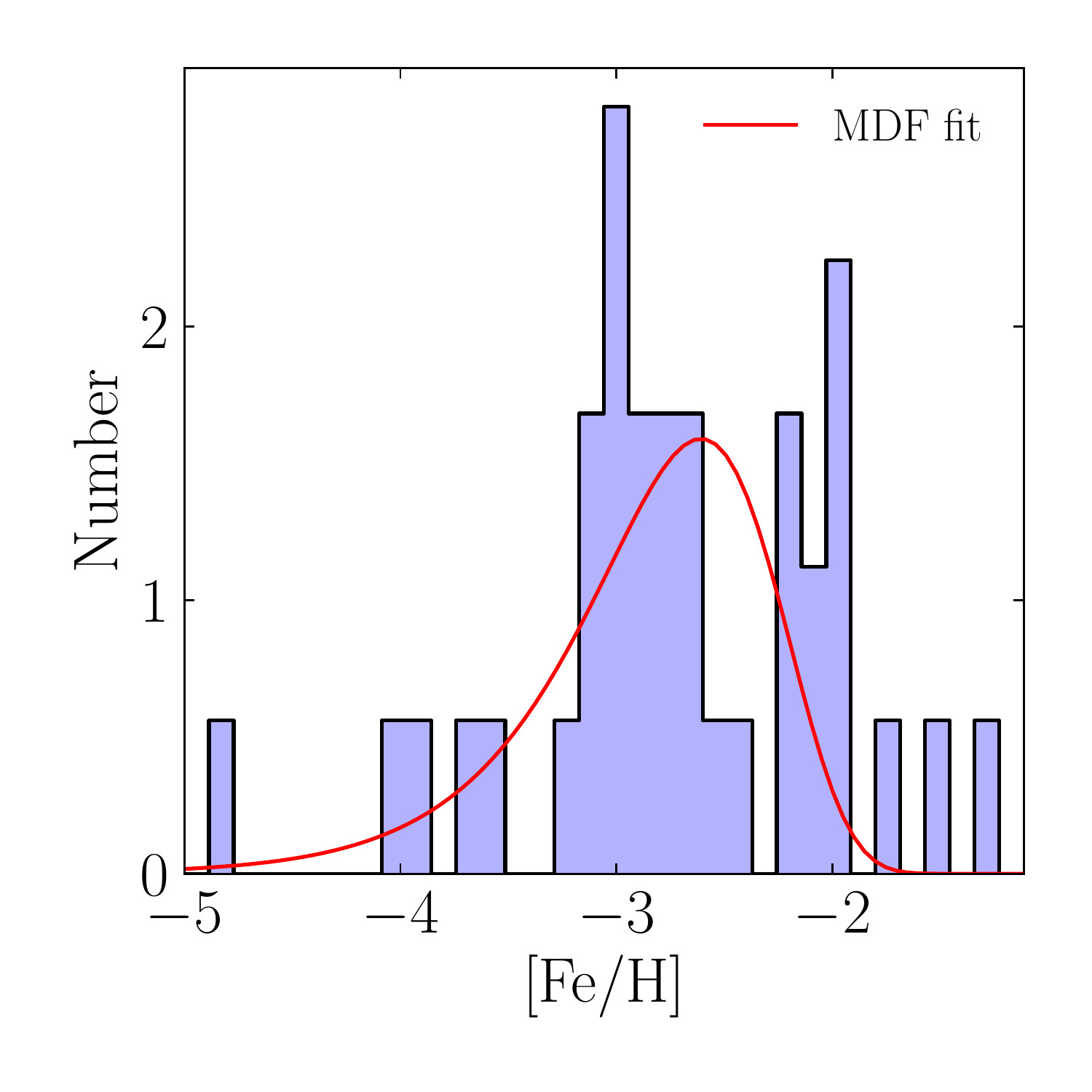} \includegraphics[width=0.3\textwidth]{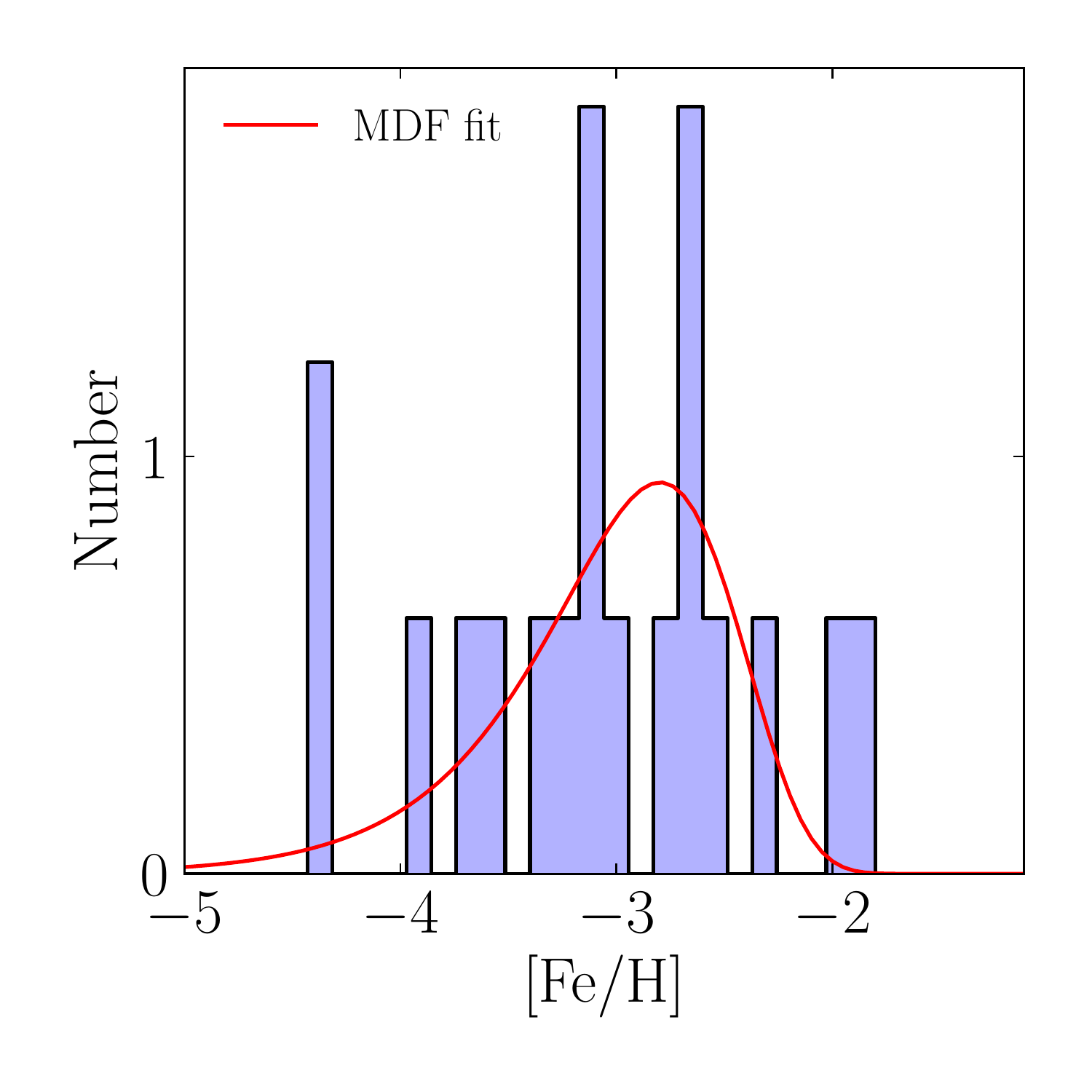}
\caption{\small 
Application of the fitting method on three  UFDs of decreasing luminosity (from left to right, $L_{\rm{V}} = 3.9 \cdot 10^4$ $\mathrm{L}_{\odot}$, $L_{\rm{V}} = 1.9 \cdot 10^4$ $\mathrm{L}_{\odot}$ and $L_{\rm{V}} = 1.0 \cdot 10^4$ $\mathrm{L}_{\odot}$).
The MDF fit given by Eq.~\ref{eq:pagel1997} is displayed in red.}
\label{fig:mdf1}
\end{figure}

We used a method where the galaxy metallicity is derived from the mode of a fit to the MDF. 
The fitting function is chosen as the MDF predicted by a 
simple chemical evolution model which assumes an instantaneous recycling \citep[see][]{Pagel1997}~:
\begin{equation} 
\frac{{\rm d}N}{{\rm d}\rm{[Fe/H]}} \propto 10^{\rm{[Fe/H]}}\exp \left ( - \frac{10^{\rm{[Fe/H]}}}{p} \right ). \label{eq:pagel1997} 
\end{equation}
Here, $p$ is a free parameter which is related to the position of the MDF mode by the relation $\text{[Fe/H]}_{\text{mode}} = p \ln(10)$. 
Fig.~\ref{fig:mdf1} illustrates the application of the method on three different dwarfs. It demonstrates
its ability in deriving a reliable mode, even if the histogram is noisy or if a large peak is found outside 
the main range of the MDF.


\end{appendix}

\end{document}